\documentstyle[11pt]{article}
   \newtheorem{Th}{Theorem}
   \newtheorem{Prop}{Proposition}
   \newtheorem{Lem}{Lemma}
   \newcommand{\be}{\begin{equation}}
   \newcommand{\ee}{\end{equation}}
   \newcommand{\lb}{\label}
   \newcommand{\lam}{\lambda}
   \newcommand{\eps}{\epsilon}
   \newcommand{\en}{\epsilon}
   \newcommand{\ba}{{\bf a}}
   \newcommand{\br}{{\bf r}}
   
   \newcommand{\hr}{{\widehat{r}}}
   \newcommand{\bv}{{\bf v}}
   \newcommand{\bk}{{\bf k}}
   \newcommand{\bx}{{\bf x}}
   \newcommand{\by}{{\bf y}}
   
   \newcommand{\bD}{{\bf D}}
   \newcommand{\bR}{{\bf R}}
   
   \newcommand{\bG}{{\bf G}}
   \newcommand{\bV}{{\bf V}}
   \newcommand{\bz}{{\bf 0}}
   
   \newcommand{\BZ}{{\bf Z}}
   \newcommand{\bxi}{{\mbox{\boldmath $\xi$}}}
   \newcommand{\Bxi}{{\mbox{\boldmath $\Xi$}}}
   
   \newcommand{\oH}{\widetilde{{\cal H}}}
   \newcommand{\cH}{{\cal H}}
   \newcommand{\cS}{{\cal S}}
   \newcommand{\of}{\overline{f}}
   
   \newcommand{\ch}{\widetilde{h}}
   \newcommand{\grad}{{\mbox{\boldmath $\nabla$}}}
   \newcommand{\bdot}{{\mbox{\boldmath $\cdot$}}}
   \newcommand{\btimes}{{\mbox{\boldmath $\times$}}}

\begin{document}

   \title{Existence and Uniqueness of $L^2$-Solutions at Zero-Diffusivity
	  in the Kraichnan Model of a Passive Scalar}

   \author{Gregory Eyink and Jack Xin\\
	   {\em Department of Mathematics}\\
	   {\em University of Arizona}}

   \date{\today}
   \maketitle
   \begin{abstract}
   We study Kraichnan's model of a turbulent scalar, passively advected by 
   a Gaussian random velocity field delta-correlated in time, for every 
   space dimension $d\geq 2$ and eddy-diffusivity (Richardson) exponent
   $0<\zeta<2$. We prove that at zero molecular diffusivity, or 
   $\kappa = 0$, there exist unique weak solutions in
   $L^2\left(\Omega^{\otimes N}\right)$ to the singular-elliptic, linear 
   PDE's for the stationary $N$-point statistical correlation functions, 
   when the scalar field is confined to a bounded domain $\Omega$ with 
   Dirichlet b.c. Under those conditions we prove that the $N$-body 
   elliptic operators in the $L^2$ spaces have purely discrete, positive 
   spectrum and a minimum eigenvalue of order $L^{-\gamma}$, with $\gamma
   =2-\zeta$ and with $L$ the diameter of $\Omega$. We also prove that the 
   weak $L^2$-limits of the stationary solutions for positive, $p$th-order 
   hyperdiffusivities $\kappa_p>0$, $p\geq 1$, exist when $\kappa_p
   \rightarrow 0$ and coincide with the unique zero-diffusivity solutions. 
   These results follow from a lower estimate on the minimum eigenvalue of 
   the $N$-particle eddy-diffusivity matrix, which is conjectured for 
   general $N$ and proved in detail for $N=2,3,4$. Some additional issues 
   are discussed: (1) H\"{o}lder regularity of the solutions; (2) the 
   reconstruction of an invariant probability measure on scalar fields 
   from the set of $N$-point correlation functions, and (3) time-dependent 
   weak solutions to the PDE's for $N$-point correlation functions with 
   $L^2$ initial data.
   \end{abstract}

   \newpage

   \section{Introduction}

   We study the model problem of a scalar field $\theta(\br,t)$ satisfying an
   advection-diffusion equation
   \be (\partial_t+\bv\bdot\grad_\br)\theta=\kappa\bigtriangleup_\br\theta+f
   \lb{pseq} \ee
   in a bounded domain $\Omega$ of Euclidean $d$-dimensional space ${\bf R}^d$,
   with Dirichlet conditions on the boundary
   $\partial\Omega$. The scalar source $f(\br,t)$ is assumed a Gaussian random
   field, white-noise in time but regular in space.
   Precisely, we take $f$ with mean $\langle f(\br,t)\rangle =\of(\br)\in
   L^2(\Omega)$ and covariance
   \be \langle f(\br,t)f(\br',t')\rangle-\langle f(\br,t)\rangle\langle
   f(\br',t')\rangle
						  =F(\br,\br')\delta(t-t')
   \lb{Fcov} \ee
   with $F\in L^2\left(\Omega\otimes\Omega\right)$. The velocity field is also
   assumed Gaussian, white-noise in time,
   zero-mean with covariance
   \be \langle v_i(\br,t)v_j(\br',t')\rangle =V_{ij}(\br-\br')\delta(t-t')
   \lb{Vcov} \ee
   The velocity to be considered is a divergence-free random field in ${\bf R}^d$
   and, for convenience, statistically
   homogeneous. There is no reason to insist on Dirichlet b.c. for the velocity
   field. The spatial covariance matrix
   $\bV$ we consider is defined by the Fourier integral
   \be V_{ij}(\br)= D_0\int {{d^d\bk}\over{(2\pi)^d}}\,\,
		 \left(k^2+m^2\right)^{-(d+\zeta)/2}P^\perp_{ij}(\bk)
       e^{i\bk\bdot\br}. \lb{Vspec} \ee
   where $0<\zeta<2$ and $P^\perp_{ij}(\bk)$ is the projection in ${\bf R}^d$ onto
   the subspace perpendicular to $\bk$.
   This automatically defines a suitable positive-definite, symmetric
   matrix-valued function, divergence-free in each index.
   The model originates in the 1968 work of R. H. Kraichnan \cite{Kr68} and has
   been the subject of recent analytical
   investigations \cite{SS,Maj,GK-L,GK,BGK,CFKL,CF}. It is not hard to show that
   \be  V_{ij}(\br)\sim V_0\delta_{ij}-D_1\cdot r^\zeta\cdot\left[\delta_{ij}
   +{{\zeta}\over{d-1}}\left(\delta_{ij}-{{r_ir_j}\over{r^2}}\right)\right]+\cdots
   \lb{scaleq} \ee
   asymptotically for $mr\ll 1$, with $V_0$ and $D_1$ constants proportional to
   $D_0$, given below. See also
   Section 4.1 of \cite{GK-L}. The exponent $\zeta$ has the physical
   interpretation of an ``eddy-diffusivity
   exponent'' analogous to the Richardson exponent $4/3$ \cite{Rich}.

   The remarkable feature of Kraichnan's model, which makes it, in a certain
   sense, ``exactly soluble'' is that $N$-th
   order correlation functions
   $\Theta_N(\br_1,...,\br_N;t)=\langle\theta(\br_1,t)
   \cdots\theta(\br_N,t)\rangle$ satisfy
   {\em closed} equations of the form
   \begin{eqnarray}
   \, & & \partial_t\Theta_N= -\oH_N^{(\kappa)}\Theta_N+\sum_n
   \of(\br_n)\Theta_{N-1}(...\widehat{\br_n}...) \cr
   \, & & \,\,\,\,\,\,\,\,\,\,\,\,\,\,\,\,\,\,\,\,\,\,\,\,\,\,\,\,+\sum_{{\rm
   pairs}\,\,\,\,\{nm\}}
   F(\br_n,\br_m)\Theta_{N-2}(...\widehat{\br_n}...\widehat{\br_m}...).
   \lb{closeq}
   \end{eqnarray}
   In this equation for the $N$-correlator only itself and lower-order correlators
   appear \cite{Kr68,SS,Maj,GK-L}.
   Here $\oH_N^{(\kappa)}$ is an elliptic partial-differential operator in
   $\Omega^{\otimes N}$ defined as
   \be \oH_N^{(\kappa)}=
   -{{1}\over{2}}\sum_{i,j=1}^d\sum_{n,m=1}^N\,\,{{\partial}\over{\partial
   x_{in}}}
	  \left[V_{ij}(\br_n-\br_m){{\partial}\over{\partial x_{jm}}}\cdot\right]
	       -\kappa\sum_{n=1}^N\bigtriangleup_{\br_n}, \lb{singell} \ee
   with Dirichlet b.c., where $x_{in}$ are Cartesian coordinates in $\left({\bf
   R}^d\right)^{\otimes N}$. However,
   the operator $\oH_N$ obtained by taking $\kappa\rightarrow 0$ is degenerate,
   i.e. it is singular-elliptic. We refer to
   $\oH_N$ as the $N$-body {\em convective operator} because it accounts for the
   effects of the velocity advection alone in
   the equation (\ref{closeq}) for $N$-point correlations. Because of the
   degeneracy for $\kappa\rightarrow 0$, the solutions
   of the parabolic equation are expected in that limit to lie only in a
   H\"{o}lder class $C^{\gamma}\left(\Omega^{\otimes N}
   \right)$ with $\gamma=2-\zeta$. As the differential operator is of
   second-order, these solutions must then be taken in a
   suitable weak sense. Despite the degeneracy, the linear operator $\oH_N$ is
   formally self-adjoint and nonnegative in
   the $L^2$ inner product of functions on $\Omega^{\otimes N}$. This suggests
   that an $L^2$-theory of weak solutions
   to Eq.(\ref{closeq}) may be appropriate. We shall develop here such a theory in
   detail. The key to the analysis of the
   $\kappa\rightarrow 0$ limiting solutions is a proof of existence and uniqueness
   directly for $\kappa=0$.

   Let us state precisely the main theorems of this work. We shall actually
   consider a somewhat more general model than
   Eq.(\ref{pseq}), namely,
   \be (\partial_t+\bv\bdot\grad_\br)\theta=
   -\kappa_p(-\bigtriangleup_\br)^p\theta+f \lb{ppseq} \ee
   with $p\geq 1$, in which $\kappa_p$ is a so-called hyperdiffusivity of order
   $p$. This allows us to establish a universality
   result concerning the independence of limits on $p$. In this case, the closed
   correlation equations (\ref{closeq})
   are still satisfied, with the operator (\ref{singell}) replaced by
   \be \oH_N^{(\kappa_p)}=
   -{{1}\over{2}}\sum_{i,j=1}^d\sum_{n,m=1}^N\,\,
   {{\partial}\over{\partial
   x_{in}}}\left[V_{ij}(\br_n-\br_m){{\partial}\over{\partial x_{jm}}}\cdot\right]
		       +\kappa_p\sum_{n=1}^N(-\bigtriangleup_{\br_n})^p.
   \lb{psingell} \ee
   Note that this operator requires higher-order Dirichlet b.c., namely, elements
   in its domain must have zero trace on the
   boundary for the first $k=[\![p-(1/2)]\!]$ derivatives. However, our first main
   result is for the solution of that
   equation directly at $\kappa_p=0$:
   \begin{Th}
   Assume that $d\geq 2$ and $0<\zeta<2$. Then, for integers $N\geq 1$, the
   equation (\ref{closeq}) at $\kappa=0$ has a unique
   stationary weak solution $\Theta_N^*$ in $L^2\left(\Omega^{\otimes N}\right)$.
   Away from the codimension-$d$ set where
   pairs of points in $\bR=(\br_1,...,\br_N)$ coincide, the solution
   $\Theta_N^*(\bR)$ is in $H^1_0\left(\Omega^{\otimes N}
   \right)$.
   \end{Th}
   This ideal zero-diffusivity solution is, in fact, the physically relevant one
   in the limits $\kappa_p\rightarrow 0$, as
   shown by our second main result:
   \begin{Th}
   Assume that $d\geq 2$ and $0<\zeta<2$, and also $p\geq 1$.

   \noindent (i) For integers $N\geq 1$, the equation (\ref{closeq}), with
   $\oH^{(\kappa)}$ generalized to $\oH^{(\kappa_p)}$,
   has a unique stationary weak solution $\Theta_N^{(\kappa_p)*}$ in
   $L^2\left(\Omega^{\otimes N}\right)$, which, in fact,
   belongs to the Sobolev space $H^p_0\left(\Omega^{\otimes N}\right)$.

   \noindent (ii) The weak-$L^2$ limit exists as $\kappa_p\rightarrow 0$ and
   $w-\lim_{\kappa_p\rightarrow 0}
   \Theta_N^{(\kappa_p)*}=\Theta_N^*$.
   \end{Th}

   \noindent To prove these results requires a spectral analysis of the $N$-body
   convective operator $\oH_N$. In fact,
   we show that this operator has pure point spectrum, using a criterion borrowed
   from a work of R. T. Lewis \cite{Lew}.
   Discreteness of the spectrum was already shown by Majda \cite{Maj} in his
   simple version of the model. For our theorems
   above, we do not really require that $\oH_N$ have a compact inverse, but merely
   a bounded inverse. To prove
   this, we require an estimate from below on the quadratic form associated to
   $\oH_N$. This is proved in two steps.
   First, for each integer $N\geq 1$ we define the $(Nd)\times(Nd)$-dimensional
   matrix $[\bG_N(\bR)]_{in,jm}=
   \langle v_i(\br_n)v_j(\br_m)\rangle,\,\,\,i,j=1,...,d,n,m=1,...,N.$ Physically,
   this is interpreted as an
   {\em N-particle eddy-diffusivity matrix}. Mathematically, it is the nonnegative
   Gramian matrix of the $Nd$ elements
   $v_i(\br_n)$ in the $L^2$ inner-product space of the random velocity field. It
   is nonsingular if and only if these
   $Nd$ elements are linearly independent. We shall prove below (Proposition 2)
   that its minimum eigenvalue obeys
   $\lambda_N^{\min}(\bR)\geq C_N [\rho(\bR)]^\zeta$, where $\rho(\bR)=\min_{n\neq
   m}|\br_n-\br_m|$, when $N=2,3,4$.
   The second step of the proof uses only this property of $\bG_N(\bR)$, which is
   conjectured to hold for all $N\geq 1$.
   As a consequence of this estimate, we prove a lower bound on the operator
   quadratic form, reminiscent of the well-known
   {\em Hardy inequality} \cite{HLP} (Theorem 330). For the operator with
   Dirichlet b.c. we may adapt a convenient proof
   of the Hardy-type inequality due also to Lewis \cite{Lew}. Unfortunately, as
   explained below, this proof does not work
   with periodic b.c. although the inequality is likely to hold there as well (for
   zero-mean functions). Lewis' argument is
   also too restrictive to permit treatment of other models with more natural b.c.
   on the velocity field. In a real turbulent
   flow with velocity field governed by the Navier-Stokes equation, the
   realizations of the velocity field would satisfy
   also Dirichlet b.c. This behavior may be mimicked with the Gaussian random
   velocity fields by taking as their covariance
   \be V_{ij}^{(\Omega)}(\br,\br')=\Delta_\Omega(\br)V_{ij}(\br-\br')
			   \Delta_\Omega(\br'), \lb{dampvel} \ee
   in which $\Delta_\Omega(\br)$ is a suitable ``wall-damping function''. It
   should be taken as some decreasing
   function of the distance to the boundary $\partial\Omega$, vanishing there as
   some power. Of course, with this choice
   of velocity covariance, a lower bound directly follows from our present work
   that $\lambda_N^{\min}(\bR)\geq
   C_N [\rho(\bR)]^\zeta[\Delta_\Omega(\bR)]^2$, where
   $\Delta_\Omega(\bR)=\min_{1\leq n\leq N}\Delta_\Omega(\br_n)$.
   While we expect the main results of this work to carry over to such models, it
   requires a different proof of the
   generalized Hardy inequality. We will return to this problem in a later work.

   Let us summarize the contents of this paper: In Section 2 we establish the
   required properties of the model velocity
   covariance and the resulting $N$-particle eddy-diffusivity matrix, in
   particular the lower bound on the minimum
   eigenvalue. In Section 3 we study the operator quadratic form, and prove its
   principal properties, such as the
   generalized Hardy inequality. Finally, in Section 4 we exploit these results to
   prove the main Theorems 1 and 2
   above. In the conclusion Section 5 we briefly discuss three other problems:
   regularity of solutions, the
   reconstruction of an invariant measure from the stationary $N$-point
   correlation functions, and time-dependent
   solutions to the parabolic PDE's for the $N$-correlators.

   \newpage

   \section{Properties of the N-Particle Eddy-Diffusivity Matrix}

   \noindent {\em (2.1) The Velocity Covariance Matrix}

   \noindent We first state and prove the regularity properties of the velocity
   covariance matrix elements
   $(V_{ij}(\br))$ that we will need for later analysis. We have made the choice
   of Eq.(\ref{Vspec})
   just for specificity. In fact, any velocity covariance with the following
   properties would suffice.

   \begin{Lem}
   The elements of velocity covariance matrix $V_{ij}(\br)$, $\br \in \bR^d$,
   are $C^{\infty}$ in $\br$ if $\br \neq 0$, and $C^{\zeta}$ near
   $\br =\bz$, with $\zeta \in (0,2)$. Moreover, there is a positive
   number $\rho_{0}$ such that if $r \in [0,\rho_{0}]$, we have the
   local expansion:
   \be
   V_{ij}(\br) = V_{0}\delta_{ij} - D_{1}\cdot r^\zeta\cdot\left[\delta_{ij}
		    +{{\zeta}\over{d-1}}\left(\delta_{ij}-
   {{r_ir_j}\over{r^2}}\right)\right]+ O\left (m^2r^2\right )
   \label{eq:A1}
   \ee
   \end{Lem}

   \noindent {\em Proof:} The matrix $V_{ij}(\br)$ can be written as
   \be V_{ij}(\br)= V(r)\delta_{ij}+\partial_i\partial_jW(r), \lb{Vrep} \ee
   where the function $V(r)$ is defined by the integral
   \be V(r)= D_0\int
   {{d^d\bk}\over{(2\pi)^d}}\,\,\left(k^2+m^2\right)^{-(d+\zeta)/2}
   e^{i\bk\bdot\br} \lb{Bespot} \ee
   and $W(r)$ is given by the (for $d=2$, principal part) integral
   \be W(r)= D_0\int
   {{d^d\bk}\over{(2\pi)^d}}\,\,\left(k^2+m^2\right)^{-(d+\zeta)/2}{{1}\over{k^2}}
					    e^{i\bk\bdot\br}, \lb{Wftn} \ee
   so that $-\bigtriangleup W=V$. The scalar function $V(r)$ is essentially just
   the standard Bessel potential kernel
   \cite{ArS}, and may thus be expressed in terms of a modified Bessel function:
   \be V(r)=D_0{{2^{1-(\zeta/2)}m^{-\zeta}}\over{(4\pi)^{d/2}
	   \Gamma\left({{d+\zeta}\over{2}}\right)}}\cdot
		     (mr)^{\zeta/2}K_{\zeta/2}(mr). \lb{Besrep} \ee
   The Hessian $\partial_i\partial_jW(r)$ of the function $W$ of magnitude
   $r=|\br|$ alone is
   \be \partial_i\partial_jW(r)=\delta_{ij} A(r)+ \hr_i\hr_j\cdot
   r{{dA}\over{dr}}(r), \lb{Hess} \ee
   with $A(r)= W'(r)/r$ and $\widehat{\br}=\br/r$. However, because ${\rm
   Tr}\left(\grad\otimes\grad W\right)= -V$,
   a Cauchy-Euler equation follows for $A(r)$:
   \be r{{dA}\over{dr}}(r)+ d\cdot A(r)= -V(r). \lb{CEeq} \ee
   Due to the rapid decay of its Fourier transform, the function $A(r)$ is
   continuous. Thus, the
   relevant solution is found to be
   \be A(r) = -r^{-d}\int_0^r \rho^{d-1} V(\rho) d\rho. \lb{AinV} \ee
   in terms of $V(r)$. Using this expression for $A(r)$, along with
   Eq.(\ref{Hess}), we thus find
   \be V_{ij}(\br)= (V(r)+A(r))\delta_{ij}-(V(r)+d\cdot A(r))\hr_i\hr_j, \lb{Gexp}
   \ee
   for $V_{ij}$ as a linear functional of $V$. If $V$ has a power-law form,
   $V(r)=B r^\xi$, then it
   is easy to calculate that
   \be V_{ij}(\br)=Br^\xi{{d-1}\over{d+\xi}}\left[\delta_{ij}
	+{{\xi}\over{d-1}}\left(\delta_{ij}-\hr_i\hr_j\right)\right].
   \lb{powfrm} \ee
   By means of the known Frobenius series expansions for the modified Bessel
   functions (e.g. \cite{AbS},
   (9.6.2),(9.6.10)), it follows that
   \be z^\nu K_\nu(z)=
   {{\Gamma(\nu)}\over{2^{1-\nu}}}-{{\Gamma(1-\nu)}\over{\nu\cdot
   2^{1+\nu}}}z^{2\nu}
				      +O\left(z^2\right). \lb{Frobser} \ee
   {}From these terms for $K_\nu(z)$ we obtain, upon substituting
   Eq.(\ref{Besrep}) into Eq.(\ref{Gexp}), the claimed
   asymptotic expression for $V_{ij}(\br)$ in Eq.(\ref{eq:A1}), with
   \be V_0= D_0{{(d-1)\Gamma\left({{\zeta}\over{2}}\right)}\over{(4\pi)^{d/2}\cdot
   d\cdot
			  \Gamma\left({{d+\zeta}\over{2}}\right)}}\cdot
   m^{-\zeta}, \lb{Vzero} \ee
   and
   \be D_1=
   D_0{{(d-1)\Gamma\left({{2-\zeta}\over{2}}\right)}\over{(4\pi)^{d/2}\cdot
   2^\zeta\cdot\zeta\cdot
			  \Gamma\left({{d+\zeta+2}\over{2}}\right)}}. \lb{Done}
   \ee
   Finally, the Bessel function $K_\nu(z)$ is analytic in the complex plane with a
   branch cut along the
   negative real axis. Thus, the stated smoothness properties of $V_{ij}$ follow.
   $\,\,\,\,\,\Box$

   \vspace{.1in}

   \noindent We shall denote the second term on the right hand side
   of (\ref{eq:A1}) as $-r^{\zeta}Q_{ij}$. Obviously,
   $(Q_{ij})$ is positive definite uniformly in $r$. We will
   denote by $\br_{nm}= \br_{n} - \br_{m}$ the vector, and
   $r_{nm}=|\br_{n} -\br_{m}|$ the scalar distance from $\br_{n}$ to
   $\br_{m}$; $V_{ij}$ the matrix elements, and
   $\bV_{nm}$ the matrix evaluated at $\br_{nm}$. We show two more lemmas.

   \begin{Lem}
   Let $\br_{i}$, $r=1,2,3$, be any three points in $R^{d}$, and
   $r_{12} \leq r_{13}$, $r_{12} \leq r_{23}$. Then
   there is a constant $\bar{C}$ depending on $\rho_{0}$ and
   $\zeta$ in Lemma 1
   but independent of $r_{12}$, $r_{13}$, and $r_{23}$ such that:
   \[ | \bV_{13} - \bV_{23} | \leq \bar{C}r_{12}\max(r_{13}^{\zeta -1},
   r_{23}^{\zeta-1}). \]
   \end{Lem}
   {\em Proof}: If $\zeta \in (1,2)$, then $\nabla \bV \in C^{\zeta -1}$,
   and so by Lemma 1:
   \[ |\bV_{13} -\bV_{23}| = | \br_{12} \cdot \nabla_{\br_{1}}
   \bV |_{\br_{\theta}}|= |\br_{12}\cdot (\nabla_{\br_{1}}\bV|_{\br_{\theta}} -
   \nabla_{\br_{1}}\bV|_{\br = 0})|, \]
   \[ \leq \bar{c}r_{12} r_{\theta}^{\zeta -1} \leq \bar{c} r_{12}
   \max(r_{12}^{\zeta -1},r_{23}^{\zeta -1}), \]
   where $\br_{\theta} =\theta \br_{1} + (1-\theta)\br_{2}$, for some
   $\theta \in (0,1)$.
   The case $\zeta =1$ is obviously true by the mean value theorem. Now if
   $\zeta \in (0,1),\; \max (r_{13},r_{23}) \geq \rho_{0},$
   then using
   $\bV \in C^1$ away from zero, we have:
   \[ | \bV_{13} -\bV_{23} | \leq \bar{c} r_{12} \leq
   \bar{c}r_{12}(m\max (r_{13}, r_{23}))^{\zeta -1} \]
   \[ \leq \bar{c}(\rho_{0},m)r_{12}
   \max (r_{13}^{\zeta -1},r_{23}^{\zeta -1}). \]
   If $\zeta \in (0,1)$, and $\max(r_{13},r_{23}) < \rho_{0}$, we employ
   local expansion to calculate for any $\bx \not = \by$:
   \[ | V_{ij}(\bx) -V_{ij}(\by) | \leq \bar{c} |(|\bx|^{\zeta} -
   |\by|^{\zeta})[\delta_{ij}+{\zeta \over (d-1)}(\delta_{ij} - {
   x_{i} x_{j} \over x^{2}})] | \]
   \[  + \bar{c}|y^{\zeta}({x_{i}x_{j}\over x^{2}} -
   {y_{i}y_{j}\over y^{2}})| \]
   \[ \leq \bar{c}\max(x^{\zeta -1},y^{\zeta -1})|\bx -\by| +
   \bar{c}y^{\zeta}\left
   |{x_{i}x_{j}y^{2} -y_{i}y_{j}x^{2} \over x^{2}y^{2}}\right |. \]
   The latter term is just:
   \[ \bar{c}y^{\zeta}|{(x_{i}x_{j}-y_{i}y_{j})y^{2} +
   y_{i}y_{j}(y^{2}-x^{2}) \over x^{2}y^{2}} | \]
   \[ = \bar{c}y^{\zeta}\left ({|\bx -\by|\over |x|} +
   {|\bx -\by|y \over x^{2}} + {|\bx -\by|(x+y)\over x^{2}}\right ). \]
   With no loss of generality, we assume that $y \leq x$; otherwise,
   we simply switch $\bx $ and $\by$. It follows that
   \[ |V_{ij}(\bx) -V_{ij}(\by)| \leq \bar{c}|\bx -\by|\max(x^{\zeta -1},
   y^{\zeta -1}) + \bar{c}|\bx -\by|x^{\zeta -1} \leq
   \bar{c}|\bx -\by|\max(x^{\zeta -1},y^{\zeta -1}). \]
   We complete the proof with $\bx = \br_{13}$, and
   $\by =\br_{23}$.

   \begin{Lem}
   Assume that $r_{12} \leq r_{34}$; $r_{13} =O(r_{14}) =O(r_{23}) =
   O(r_{24})$; ${r_{34} \over r_{13}} \leq \eps \in (0,1)$.
   Then there exist $\eps_{0}$ and  a positive constant $\bar{c}_{1}$
   depending on $\rho_{0}$, $\zeta$, maximum and minimum ratios of
   $r_{13}$, $r_{14}$, $r_{23}$, and $r_{24}$, such that:
   \[ |\bV_{13} -\bV_{14} - (\bV_{23} -\bV_{24})| \leq \bar{c}_{1}r_{12}
   r_{34}r_{13}^{\zeta -2}, \]
   for all $\eps \in (0,\eps_{0})$.
   \end{Lem}
   {\em Proof}: Applying the mean value theorem to $F(\br_1) \equiv
   \bV_{13}- \bV_{14}$, we get for
   $\br_{\theta}=\theta \br_{1} + (1-\theta)\br_{2}$
   that:
   \[ F(\br_{1}) -F(\br_{2}) = \br_{12} \cdot\nabla_{\br_{1}}F|_{\br_{\theta}}.\]
   If $\max (r_{13},r_{24}) \geq {\rho_{0}\over 2}$, then
   \[ \nabla_{\br_{1}}F|_{\br_{\theta}} =\nabla_{\br_{1}}\bV_{13}
   -\nabla_{\br_{1}}\bV_{14}|_{\br_{1} =\br_{\theta}}. \]
   By the smoothness of $\nabla_{\br_1}\bV_{1i}$ when the
   distance of $\br_1$ from $\br_{i},\,\,\,i=3,4$ is larger than
   ${\rho_{0} \over 4}$ (which is possible if $\eps$ is small enough),
   we obtain:
   \[ | \nabla_{\br_{1}}F|_{\br =\br_{\theta}} | \leq \bar{c}_{1}\rho_{0}r_{34},
   \]
   from which it follows that:
   \[ |F(\br_{1}) -F(\br_{2})| \leq \bar{c}_{1}r_{12} r_{34} \leq
   \bar{c}_{1}r_{12}r_{34}r_{13}^{\zeta -2}. \]
   On the other hand, if $\max(r_{13},r_{24}) < {\rho_{0} \over 2}$, we
   use local expansion in Lemma 1 to get for each matrix element:
   \begin{eqnarray}
   (F(\br_{1}) - F(\br_{3}))_{ij} & = & (-D_{1})r_{13}^{\zeta}[
   \delta_{ij} + {\zeta \over d-1}(\delta_{ij} -{\br_{13}^{(i)}\br_{13}^{(j)}
   \over r_{13}^{2}})] \nonumber \\
   & + & D_{1} r_{14}^{\zeta}[\delta_{ij}+{\zeta \over d-1}(
   \delta_{ij} -{\br_{14}^{(i)}\br_{14}^{(j)} \over r_{14}^{2}})] \nonumber \\
   & - & (1 \rightarrow 2) \nonumber \\
   & = & (-D_{1})\br_{12} \cdot \nabla_{\br_{1}}(
   r_{13}^{\zeta}[
   \delta_{ij} + {\zeta \over d-1}(\delta_{ij} -{\br_{13}^{(i)}\br_{13}^{(j)}
   \over r_{13}^{2}})] \nonumber \\
   & - & r_{14}^{\zeta}[\delta_{ij}+{\zeta \over d-1}(
   \delta_{ij} -{\br_{14}^{(i)}\br_{14}^{(j)} \over r_{14}^{2}})])
   (\br_{1} =\br_{\theta}), \label{eq:L1}
   \end{eqnarray}
   where the notation $(1 \rightarrow 2)$ means the same terms as
   before except that subscript $1$ is replaced by $2$.
   Let us calculate the $\br_{1}$ gradient in (\ref{eq:L1}) as ($k$ meaning
   the $k$th component of this gradient):
   \[
   \zeta r_{13}^{\zeta -1}{\br_{1}^{(k)} -\br_{3}^{(k)} \over
   r_{13}}
   [\delta_{ij}+{\zeta \over d-1}(
   \delta_{ij} -{\br_{13}^{(i)}\br_{13}^{(j)} \over r_{13}^{2}})]
    + r_{13}^{\zeta}{-\zeta\over d-1}\cdot \nabla_{\br_{1}}
   {\br_{13}^{(i)}\br_{13}^{(j)}\over r_{13}^{2}} - (3\rightarrow 4) \]
   \begin{eqnarray}
   & =& \zeta (r_{13}^{\zeta -1}-r_{14}^{\zeta -1})
   {\br_{1}^{(k)} -\br_{3}^{(k)} \over
   r_{13}}
   [\delta_{ij}+{\zeta \over d-1}(
   \delta_{ij} -{\br_{13}^{(i)}\br_{13}^{(j)} \over r_{13}^{2}})]
   \nonumber \\
   & + & \zeta r_{14}^{\zeta -1}\left (
   {\br_{1}^{(k)} -\br_{3}^{(k)} \over
   r_{13}}
   [\delta_{ij}+{\zeta \over d-1}(
   \delta_{ij} -{\br_{13}^{(i)}\br_{13}^{(j)} \over r_{13}^{2}})] -
   (3 \rightarrow 4) \right ) \nonumber \\
   & + & {-\zeta \over d-1}\left (
   r^{\zeta}_{13}{-2\br_{13}^{(i)}\br_{13}^{(j)}\br_{13}^{(k)}\over r_{13}^{4}}
   +r^{\zeta}_{13}{\delta_{ik}\br_{13}^{(j)}\over r_{13}^{2}} +
   r_{13}^{\zeta}{\br_{13}^{(i)}\delta_{jk}\over r_{13}^{2}} -
   (3 \rightarrow 4) \right ). \label{eq:L2}
   \end{eqnarray}
   Note that the first term of the right hand side of (\ref{eq:L2})
   is bounded by:
   \[ C(\zeta,d)|r_{13}^{\zeta -1} -r_{14}^{\zeta -1}| \leq C(\zeta,d)
   r_{23}^{\zeta -2}r_{34}. \]
   We can think of
   \[
   {\br_{1}^{(k)} -\br_{3}^{(k)} \over
   r_{13}}
   [\delta_{ij}+{\zeta \over d-1}(
   \delta_{ij} -{\br_{13}^{(i)}\br_{13}^{(j)} \over r_{13}^{2}})]  \]
   as a bounded $C^{1}$ function of the unit vector
   $\hat{\br}_{13}$ along $\br_{13}$. Hence the second term
   of (\ref{eq:L2})
   being the difference of two values of this function at two points
   $\hat{\br}_{13}$ and $\hat{\br}_{14}$ is of the order
   $O({r_{34}\over r_{13}})$. Thus the second term is bounded by
   \[ \bar{c}_{1}r_{14}^{\zeta -1}r_{34}r_{13}^{-1}\leq
   \bar{c}_{1}r_{13}^{\zeta -2}r_{34}.\]
    Similarly, the third term is
   bounded as such. Combining the above with (\ref{eq:L1}) we deduce that
   $|F(\br_{1})-F(\br_{3})| \leq \bar{c}_{1}r_{12}r_{34}r_{13}^{\zeta -2}$.
   The proof of the lemma is complete.

   \noindent {\em (2.2) The N-Point Eddy-Diffusivity (Gramian) Matrix}

   \noindent As in the Introduction, we define for each integer $N\geq 1$ the
   $(Nd)\times(Nd)$-dimensional Gramian matrix
   $[\bG_N(\bR)]_{in,jm}=\langle v_i(\br_n)v_j(\br_m)\rangle.$ For the moment we
   consider general velocity covariances,
   given by a Fourier integral
   \be V_{ij}(\br)= \int
   {{d^d\bk}\over{(2\pi)^d}}\,\,\widehat{V}_{ij}(\bk)e^{i\bk\bdot\br}, \lb{gVspec}
   \ee
   with $\widehat{\bV}(\bk)\geq \bz$ for each $\bk\in\bR^d$. The basic properties
   are contained in:
   \begin{Prop}For each $N\geq 2$ the matrix $\bG_N(\bR)$ has the following
   properties:

   (i) $\bG_N(\bR)\geq \bz$.

   (ii) Assume that for all $\bk\in\bR^d$ the velocity spectral matrix
   $\widehat{\bV}(\bk)>\bz$ on the subspace
   orthogonal to the vector $\bk$. In that case, $\bG_N(\bR)$ has a nontrivial
   null space if and only if
   $\br_n=\br_m$ for some pair of points $n\neq m$.

   (iii) For the same hypothesis as (ii), if $\{\br_1,...,\br_N\}$ has $K$ subsets
   of coinciding points, with
   $N_k$ points in the $k$th subset, $k=1,...,K,$ then the dimension of the null
   space of $\bG_N(\bR)$ is $\sum_{k=1}^K
   (N_k-1)d.$ The null space consists precisely of vectors
   $\Bxi=(\bxi_1,...,\bxi_N)$ with the property that
   \be \sum_{n_k=1}^{N_k} \bxi_{n_k}=\bz, \lb{kernel} \ee
   for each $k=1,...,K$, where the sum runs over the $N_k$ coinciding points in
   the $k$th subset.
   \end{Prop}
   {\em Proof}: {\em (i)} Obvious from the stochastic representation. {\em (ii)}\&
   {\em (iii)}
   Let us assume that the $Nd$-dimensional vector $\Bxi=(\bxi_1,...,\bxi_N)$
   belongs to ${\rm Ker}\bG_N(\bR)$. Then, using the
   definition of $\bG_N(\bR)$ and the Fourier integral representation
   Eq.(\ref{gVspec}), it follows that
   \be 0=\langle\Bxi,\bG_N(\bR)\Bxi\rangle=
	     \int {{d^d\bk}\over{(2\pi)^d}}\,\,\overline{\left(\sum_{n=1}^N\bxi_n
   e^{i\bk\bdot\br_n}\right)}\bdot
	     \widehat{\bV}(\bk)\bdot\left(\sum_{n=1}^N\bxi_n
   e^{i\bk\bdot\br_n}\right). \lb{nullcon1} \ee
   This can only occur if the nonnegative integrand vanishes for a.e.
   $\bk\in\bR^d$. Because of our assumption
   on $\widehat{\bV}(\bk)$, this implies that
   \be \sum_{n=1}^N\bxi_n e^{i\bk\bdot\br_n}=\alpha(\bk)\cdot\bk, \lb{nullcon2}
   \ee
   for a.e. $\bk\in\bR^d$ with some complex coefficient $\alpha(\bk)$. Taking the
   vector cross product with respect to
   $\bk$ and then Fourier transforming, we obtain that
   \be  \sum_{n=1}^N \bxi_n\btimes \grad\delta(\br-\br_n)=\bz, \lb{nullcon3} \ee
   in the sense of distributions. Therefore, for any smooth test function
   $\varphi$,
   \be  \sum_{n=1}^N \bxi_n\btimes (\grad\varphi)(\br_n)=\bz. \lb{nullcon4} \ee
   Because the values of $\grad\varphi$ may be arbitrarily specified at any set of
   distinct points, it follows that
   \be \sum_{k=1}^K \left(\sum_{n_k=1}^{N_k}\bxi_{n_k}\right)\btimes \ba_k=\bz
   \lb{nullcon5} \ee
   with $\ba_k\in\bR^d$ arbitrary. This immediately implies that Eq.(\ref{kernel})
   is both necessary and sufficient for
   $\Bxi$ to belong to ${\rm Ker}\bG_N(\bR)$. Furthermore, this subspace has
   dimension $\sum_{k=1}^K (N_k-1)d$, which
   completes the proof of {\em (iii)}. Finally, {\em (ii)} follows from {\em
   (iii)} by observing that ${\rm Ker}\bG_N(\bR)
   =0$ if and only if $K=N$ and $N_k=1$ for all $k=1,...,N$. $\,\,\,\,\,\Box$

   For the particular choice of covariance function defined by Eq.(\ref{Vspec})
   for $0<\zeta<2$, we need also
   the following crucial lower bound:
   \begin{Prop} For each $0<\zeta<2$ and $d\geq 2$, there exists for each $N\geq
   2$ a constant $C_N=C_N(d,\zeta)>0$ so that
   the minimum eigenvalue $\lambda_N^{\min}(\bR)$ of $\bG_N(\bR)$ satisfies
   \be \lambda_N^{\min}(\bR)\geq C_N\cdot [\rho(\bR)]^\zeta, \lb{lowbd} \ee
   with $\rho(\bR)=\min_{n\neq m}r_{nm}$.
   \end{Prop}
   \noindent The above property will be proved in detail in this paper for
   $N=2,3,4$. While the proof in these cases strongly
   suggests the result is true for all $N\geq 2$, the argument becomes
   increasingly complicated for larger values
   of $N$. We shall leave the discussion of the general $N$ to a future
   publication, although we point out that many parts
   of the argument below apply for the general case. Note that we can view $\bG_N$
   as a matrix parametrized by the
   $\zeta$ power of the minimum distance, $\eps \equiv \rho^{\zeta}$. Let
   $\lambda_N^{\min}=\lambda_N(\eps)$ be the
   minimum positive eigenvalue of $\bG_N$ with corresponding unit eigenvector
   $\Bxi_N^{\min}=\Bxi_N(\eps)$. Then,
   by the standard formulae of degenerate first-order perturbation theory (see
   Kato, \cite{Kato}):
   \be
   \lam_N(\eps)  = \langle\Bxi_N(0),\bG_N(\eps)\Bxi_N(0)\rangle+ O(\eps^{2}).
   \label{eq:A3}
   \ee
   We have used the fact that $\lam_N(\eps)$ is at least twice differentiable in
   $\eps$ near zero: see \cite{Kato},
   Theorems II.1.8 and II.6.8. Furthermore, $\Bxi_N(0)$ is in the null space of
   $\bG_N(0)$. Thus, by Proposition 1{\em (iii)},
   $\Bxi_N(0)=(\bxi_{1},...,\bxi_{N})$ such that $\sum_{n=1}^N\bxi_{n} = 0$. By
   simply minimizing over this entire subspace
   of vectors $\Bxi$, we shall show that the righthand side quadratic form of
   (\ref{eq:A3}), denoted by $Q_N(\bxi_{1},\cdots,
   \bxi_{N-1})$, is bounded from below by a constant times $\eps$. Thus
   $\lam(\eps)$ obeys the same type of lower bound.

   \noindent {\bf Proposition 2, N=3 Case}

   {\bf \noindent Remark:} The following proof for Proposition 2, $N=3$,
   also implies the lower bound $C_2r_{12}^{\zeta}$ for the $N=2$ case.

   \noindent {\em Proof:} Let $\br_{n}$, $n=1,2,3$, be three distinct points in
   $\bR^{d}$, $d\geq 2$.
   Then we show that there is a positive constant $C_3= C_3(\rho_{0})$, where
   $\rho_{0}$ is the scale of local approximation (\ref{eq:A1}), such that
   the minimum eigenvalue of $\bG_3$ is bounded from below by
   $C_3 \rho^{\zeta}$. It suffices to treat the situation where $\rho \leq
   \rho_{0}$,
   otherwise, we conclude with Proposition 1. Let $C_{0}$ be a large but $O(1)$
   constant to be determined, and
   let $r_{12}=\rho$ for definiteness.

   \noindent Case I:
   Suppose now that ${ r_{13} \over \rho}\leq C_{0}$, and
   ${r_{23} \over \rho} \leq C_{0}$. By further reducing the size of
   $\rho$, we can ensure that $\rho C_{0} \leq \rho_{0}$. Now
   write:
   \[ \left ( \begin{array}{r}
	   \bxi_{1} \\
	   \bxi_{2} \\
	   -\bxi_{1} -\bxi_{2}
	   \end{array} \right )
	   = \left ( \begin{array}{rrr}
		   1 & 0 &  0 \\
		   0 & 1 & 0 \\
		   -1 & -1 & 1
		   \end{array}
	   \right ) \cdot
   \left ( \begin{array}{r}
	   \bxi_{1} \\
	   \bxi_{2} \\ 0
   \end{array} \right ), \]
   then:
   \[ Q_3 = \langle(\bxi_{1},\bxi_{2}); \left ( \begin{array}{rr}
	   2(\bV(0) -\bV_{13}) & \bV(0)+\bV_{12} -\bV_{13} -\bV_{23} \\
	   \bV(0) +\bV_{12} -\bV_{13} -\bV_{23} & 2(\bV(0) -\bV_{23})
	   \end{array} \right ) \cdot
	   \left ( \begin{array}{rr}
		   \bxi_{1} \\\bxi_{2}
	   \end{array} \right ) \rangle. \]
   Since all the three distances are less than $\rho_{0}$, we
   apply lemma 1 to see that
   ${|\bV(0) -\bV_{ij}|\over r_{ij}^{\zeta}} \leq C_{0}$.
   Therefore we can factor out $\rho^{\zeta}$. The remaining
   entries are bounded by $C_{0}$, and we also know that
   they form a positive definite matrix. Hence by
   continuity of eigenvalues on the matrix entries, we get the bound:
   \be
   Q_3 \geq \mu_{1}(C_{0})\rho^{\zeta}, \label{eq:A4}
   \ee
   for some positive constant $\mu_{1}=\mu_{1}(C_{0})$.

   \noindent Case II: Suppose ${r_{13}\over \rho} > {C_{0} \over 2}$,
   ${r_{23} \over \rho} > { C_{0}\over 2}$. By geometric
   constraint, $\frac{r_{13}}{r_{23}} = 1 + O(C_{0}^{-1})$. To
   estimate $Q_3$ from below, we decompose the vectors
   $\{ (\bxi_1,\bxi_2,-(\bxi_1 + \bxi_2)) \}$ into the orthogonal sum of
   $\{ (\bar{\bxi}_{1},-\bar{\bxi}_{1},0)\}$ and
   $\{(\bxi'_1,\bxi'_1,-2\bxi'_1)\}$. Then $Q_3$ is expressed into the sum of
   three terms as:
   \[ Q_3(\bxi_1,\bxi_2) = \langle (\bxi_1,\bxi_2,-(\bxi_1 + \bxi_2)), \bG_3
   (\bxi_1,\bxi_2,-(\bxi_1 +\bxi_2))^{T}\rangle
   \]
   \[ = \langle(\bar{\bxi}_{1},-\bar{\bxi}_{1},0),
   \bG_3(\bar{\bxi}_{1},-\bar{\bxi}_{1},0)^{T}\rangle \]
   \[ + \langle
   (\bxi'_{1},\bxi'_1,-2\bxi'_1),\bG_3(\bxi'_1,\bxi'_1,-2\bxi'_1)^{T}\rangle \]
   \be +
   2\langle(\bar{\bxi}_1,-\bar{\bxi}_{1},0),
   \bG_3(\bxi'_1,\bxi'_1,-2\bxi'_1)^{T}\rangle.
   \label{eq:A5}
   \ee
   Write:
   \[ \left (\begin{array}{r}
   \bar{\bxi}_{1}\\ - \bar{\bxi}_{1}\\ 0
   \end{array} \right ) =
   \left (\begin{array}{rrr}
   1 & 0 & 0 \\
   -1 & 1 & 0 \\
   0 & 0 & 1
   \end{array}
   \right ) \left (\begin{array}{r}
   \bar{\bxi}_1\\ 0\\ 0
   \end{array} \right ), \]
   then the bar term of (\ref{eq:A5}):
   \[ \langle(\bar{\bxi}_{1},0,0);
   \left (\begin{array}{rrr}
   1 & -1 & 0 \\
   0 & 1 & 0 \\
   0 & 0 & 1
   \end{array}
   \right )\left (\begin{array}{rrr}
   \bV(0)-\bV_{12} & \bV_{12} & \bV_{13}\\
   \bV_{12} -\bV(0) & \bV(0) & \bV_{23} \\
   \bV_{13}-\bV_{23} & \bV_{23} & \bV(0)
   \end{array}
   \right )\left ( \begin{array}{r}
   \bar{\bxi}_{1} \\0 \\0
   \end{array} \right ) \rangle  \]
   \be
    = 2 \langle\bar{\bxi}_1,(\bV(0)-\bV_{12})\bar{\bxi}_1\rangle \geq
   \bar{c}_1\rho^{\zeta}
   |\bar{\bxi}_1|^{2},
   \label{eq:A6}
   \ee
   where $\bar{c}$ here and after will denote a positive constant
   depending only on $\rho_{0}$. Also $1$ is a shorthand
   for $d\times d$ identity matrix. Similarly, we express:
   \[ \left (\begin{array}{r}
   \bxi'_1 \\ \bxi'_{1}\\ -2\bxi'_{1} \end{array}
   \right ) = \left (\begin{array}{rrr}
   1 & 0 & 0 \\
   1 & 1 & 0 \\
   -2 & 0 & 1
   \end{array}
   \right )\left (\begin{array}{r}
   \bxi'_1 \\ 0 \\ 0 \end{array}
   \right ) \]
   and write the prime term by Lemma 2 as:
   \[ \langle(\bxi'_{1},0,0), \left (\begin{array}{rrr}
   1 & 1 & -2 \\
   0 & 1 & 0 \\
   0 & 0 & 1
   \end{array} \right )\left (\begin{array}{rrr}
   \bV(0)+\bV_{12}-2\bV_{13} & \bV_{12} & \bV_{13}\\
   \bV_{12} +\bV(0)-2\bV_{23} & \bV(0) & \bV_{23} \\
   \bV_{13}+\bV_{23}-2\bV(0) & \bV_{23} & \bV(0)
   \end{array}
   \right )\left (\begin{array}{r}
   \bxi'_1 \\ 0 \\ 0 \end{array}
   \right )\rangle\]
   \[ = \langle \bxi'_1,(6\bV(0)+2\bV_{12} -4\bV_{13}-4\bV_{23})\bxi'_1\rangle \]
   \[ = \langle\bxi'_1,8(\bV(0)-\bV_{13})\bxi'_1\rangle +
   \langle\bxi'_1,\left(2(\bV_{12}-\bV(0)) +
   4(\bV_{13}-\bV_{23})\right )\bxi'_{1}\rangle\]
   \[
   \geq \bar{c}r_{13}^{\zeta}|\bxi'_{1}|^{2} - \bar{c}(\rho^{\zeta} +
   r_{12}r_{13}^{\zeta -1})|\bxi'_{1}|^{2} \]
   \be
   \geq \bar{c}r_{13}^{\zeta}|\bxi'_{1}|^{2}( 1-\bar{c}\left(
   (\rho r_{13}^{-1})^{\zeta} + (\rho r_{13}^{-1})\right ))\geq \bar{c}_1
   r^{\zeta}_{13}
   |\bxi_1'|^{2}.
   \label{eq:A7}
   \ee
   The mixed term is equal to :
   \[ \langle (\bar{\bxi}_1,-\bar{\bxi}_{1},0), \left (\begin{array}{rrr}
   \bV(0)+\bV_{12}-2\bV_{13} & \bV_{12} & \bV_{13}\\
   \bV_{12} +\bV(0)-2\bV_{23} & \bV(0) & \bV_{23} \\
   \bV_{13}+\bV_{23}-2\bV(0) & \bV_{23} & \bV(0)
   \end{array}
   \right )\left (\begin{array}{r}
   \bxi'_1 \\ 0 \\ 0 \end{array}
   \right )\rangle\]
   \[ = \langle\bar{\bxi}_1,(\bV(0) +\bV_{12} -2\bV_{13})\bxi'_{1}\rangle -
   \langle\bar{\bxi}_{1}, (\bV_{12}+\bV(0)-2\bV_{23})\bxi'_{1}\rangle, \]
   \[ = \langle\bar{\bxi}_1,2(\bV_{23}-\bV_{13})\bxi'_{1}\rangle, \]
   and so is bounded by:
   \be
   | mixed\;\; term| \leq \bar{c}_2 |\bxi'_1|\cdot |\bar{\bxi}_{1}|\cdot r_{12}
   r_{13}^{\zeta -1}. \label{eq:A8}
   \ee
   Thus:
   \be
   Q_3 = Q_3(\bxi_1,\bxi_2) \geq \bar{c}_1\rho^{\zeta}|\bar{\bxi}_{1}|^{2}
   + \bar{c}_1 r_{13}^{\zeta}|\bxi'_{1}|^{2}
   - \bar{c}_2 |\bxi'_1|\cdot |\bar{\bxi}_{1}|r_{12}r_{13}^{\zeta -1}
   \label{eq:A9}
   \ee
   The mixed term may then be controlled by the positive terms through the
   following
   Young's inequality:
   \begin{eqnarray}
   |\bxi'_1|\cdot |\bar{\bxi}_{1}|\rho r_{13}^{\zeta -1}
	      & = & \sqrt{\theta\rho^\zeta}|\bar{\bxi}_{1}|\cdot
		    {{\rho^{1-{{\zeta}\over{2}}}r_{13}^{\zeta
   -1}}\over{\sqrt{\theta}}}|\bxi'_1| \cr
	  \,  & \leq & {{1}\over{2}}\theta\cdot \rho^\zeta|\bar{\bxi}_{1}|^2+
		       {{(\rho/r_{13})^{2-\zeta}}\over{2\theta}}\cdot
   r_{13}^{\zeta}|\bxi'_1|^2,
   \label{eq:A10}
   \end{eqnarray}
   with $\theta$ a small number in $(0,1)$. Then, since $\rho/r_{13}<2C_0^{-1}$,
   it follows that
   for any $\zeta<2$, $(\rho/r_{13})^{2-\zeta}<\theta^2$ for $C_0$ large enough.
   Thus,
   \be |\bxi'_1|\cdot |\bar{\bxi}_{1}|\rho r_{13}^{\zeta -1}
	      \leq {{1}\over{2}}\theta\cdot \rho^\zeta|\bar{\bxi}_{1}|^2+
		   {{1}\over{2}}\theta\cdot r_{13}^{\zeta}|\bxi'_1|^2,
   \label{eq:A11} \ee
   which allows the mixed term to be absorbed into the positive bar and prime
   terms.
   Combining (\ref{eq:A9}-\ref{eq:A11}), we conclude that:
   \be
   Q_3(\bxi_1,\bxi_2) \geq \bar{c}\rho^{\zeta}|\bar{\bxi}_{1}|^{2} +
   \bar{c}r_{13}^{\zeta}|\bxi'_1|^{2}, \label{eq:A12}
   \ee
   which in the original $(\bxi_1,\bxi_2)$ variables reads:
   \be
   Q_3(\bxi_1,\bxi_2) \geq \bar{c}\rho^{\zeta}|\bxi_1 -\bxi_2|^{2} + \bar{c}
   r_{13}^{\zeta}|\bxi_1 + \bxi_2|^{2}. \label{eq:A13}
   \ee
   We finish the proof with inequality (\ref{eq:A13}) and (\ref{eq:A4}).
   $\,\,\,\,\,\Box$

   \noindent {\bf Proposition 2, N=4 Case}

   \noindent We now turn to $N=4$, for which inequality (\ref{eq:A13})
   is very helpful. Let $\br_{n}$, $n=1,2,3,4$, be four distinct points in
   $\bR^d$, $d\geq 2$, and assume that $r_{12}$
   is the minimum length $\rho$. Then we show that there is a positive constant
   $\bar{c}$ depending only
   on $\rho_0$ so that the minimum eigenvalue of $\bG_4$ is bounded from below by
   $\bar{c}\rho^{\zeta}$.

   \noindent {\em Proof:} We order $r_3$ and $r_4$ according to the lengths of the
   three
   sides intersecting at them. The longest length at $r_4$ is larger
   than that at $r_3$. If they are equal, then the second longest
   length at $r_4$ is larger than its counterpart at $r_3$, and so on.
   Generically, we are able to order $r_3$ and $r_4$ this way.
   Now $r_i$, $i=1,2,3,4$, determine a tetrahedra in $R^d$. Due to
   geometric constraint, $r_{23}$ and $r_{13}$ are on the same order.
   So are $r_{14}$ and $r_{24}$. With no loss of generality, we can assume that
   $r_{13}=r_{23} =\alpha$, and $r_{14} =r_{24} =\beta$. Let
   $r_{34}$ be $\gamma$, which satisfies the inequalities:
   \be
   \gamma \leq \alpha +\beta, \; \beta \leq \alpha + \gamma; \alpha
   \leq \beta. \label{eq:A14}
   \ee
   We consider all the possibilities under (\ref{eq:A14}).

   \noindent Case I. Suppose $2 \geq {\gamma \over \beta} \geq C_{2}^{-1}$,
   where $C_2 >0$ is a large constant to be selected. We have
   four subcases: I 1.1: $ 1 \leq {\beta \over \alpha}\leq C_1$ and
   $1\leq {\alpha \over \rho}\leq C_0$;
   I 1.2: $1 \leq {\beta \over \alpha}\leq C_1$ and
   ${\alpha \over \rho} > C_0$; I 2.1: ${\beta \over \alpha} > C_1$ and
   $1\leq {\alpha \over \rho}\leq C'_0$; I 2.2:
   ${\beta \over \alpha} > C_1$ and
   $1\leq {\alpha \over \rho}>  C'_0$. Case II: $ {\gamma \over \beta}
   < C_{2}^{-1}$, which implies with (\ref{eq:A14}) that
   $1\leq {\beta \over\alpha}\leq {C_2 \over C_{2} -1}$. We have
   two subcases: II 1.1: $1\leq {\alpha \over \rho}\leq C_0$ and
   II 1.2: $1\leq {\alpha \over \rho}>C_0$

   \noindent As in the analysis for $N=3$, we assume that $\rho$ is
   smaller than $\rho_0$. The I1.1 is very similar to the first case of
   $N=3$, in that all lengths are comparable to each other. Writing
   \[ (\bxi_1,\bxi_2,\bxi_3,-(\bxi_1+\bxi_2+\bxi_3)) =
   \left(\begin{array}{rrrr}
   1 & 0 & 0 & 0\\ 0 & 1 & 0 & 0\\ 0 & 0 & 1 & 0 \\-1 & -1 & -1 & 0
   \end{array} \right )(\bxi_1,\bxi_2,\bxi_3,0)^{T}, \]
   then:
   \[
   Q_4(\bxi_1,\bxi_2,\bxi_3)=
   (\bxi_1,\bxi_2,\bxi_3) \]
   \[ \left
   ( \begin{array}{rrr}
   2(\bV(0)-\bV_{14}) & \bV(0) + \bV_{12} -\bV_{14} -\bV_{24} &
   \bV(0) + \bV_{13} -\bV_{14} -\bV_{24} \\
   \bV(0) + \bV_{12} -\bV_{14} -\bV_{24} & 2(\bV(0)-\bV_{24}) &
   \bV(0)+\bV_{23}-\bV_{24}-\bV_{34}\\
   \bV(0) + \bV_{13} -\bV_{14} -\bV_{24} & \bV(0)+\bV_{23}-\bV_{24}-\bV_{34} &
   2(\bV(0)-\bV_{34}) \end{array}\right )
   \left( \begin{array}{r}
   \bxi_1\\\bxi_2 \\\bxi_3
   \end{array}\right ). \]
   Using lemma 1 again, we can factor out
   $\rho^{\zeta}$ with remaining matrix being positive and bounded. We
   find that there is $\mu=\mu(C_0,C_1,C_2)$ such that:
   \be
   Q_4 \geq \mu \rho^{\zeta}. \label{eq:A15}
   \ee

   \noindent Now for I 1.2, we decompose
   $\{(\bxi_1,\bxi_2,\bxi_3,-(\bxi_1+\bxi_2+\bxi_3))\}$
   into the orthogonal sum of
   \[ \{(\bar{\bxi}_1, -\bar{\bxi}_1, 0,0)\}\]
    and
   \[ \{(\bxi'_1,\bxi'_1,\bxi'_2,-2\bxi'_1-\bxi'_2)\}.\]  Then:
   \[
   Q_4(\bxi_1,\bxi_2,\bxi_3) =
   \langle(\bar{\bxi}_1,-\bar{\bxi}_1,0,0),
   \bG_4(\bar{\bxi}_1,-\bar{\bxi}_1,0,0)^{T}\rangle
   \]
   \[ +\langle(\bxi'_1,\bxi'_1,\bxi'_2,-2\bxi'_1-\bxi'_2),
   \bG_4(\bxi'_1,\bxi'_1,\bxi'_2,-2\bxi'_1-\bxi'_2)^{T}\rangle\]
   \be
   + 2 \langle(\bar{\bxi}_1,-\bar{\bxi}_1,0,0),
   \bG_4(\bxi'_1,\bxi'_1,\bxi'_2,-2\bxi'_1-\bxi'_2)^{T}\rangle.
   \label{eq:A16}
   \ee
   Writing:
   \[ \left ( \begin{array}{r}
   \bar{\bxi}_1 \\ -\bar{\bxi}_{1} \\ 0 \\ 0
   \end{array} \right ) = \left (
   \begin{array}{rrrr}
   1 & 1 & 0 & 0 \\ -1 & 1 & 0 & 0 \\
   0 & 0& 1 & 0 \\ 0 & 0 & 0 & 1
   \end{array}\right )\left ( \begin{array}{r}
   \bar{\bxi}_1 \\ 0 \\ 0 \\ 0
   \end{array} \right ), \]
   we see that the  bar term is equal to:
   \be
   \langle\bar{\bxi}_{1},2(\bV(0)-\bV_{12})\bar{\bxi}_{1}\rangle \geq
   \bar{c}\rho^{\zeta}
   |\bar{\bxi}_{1}|^{2}, \label{eq:A17}
   \ee
   Writing:
   \[ \left ( \begin{array}{r}
   \bxi'_1 \\ \bxi'_{1} \\ \bxi'_2 \\ -2\bxi'_1 -\bxi'_2
   \end{array} \right ) = \left (
   \begin{array}{rrrr}
   1 & 0 & 1 & 0 \\ 1 & 0 & 0 & 0 \\
   0 &  1 & 1 & 0 \\ -2 & -1 & 0 &  1
   \end{array}\right )\left ( \begin{array}{r}
   \bxi_1' \\ \bxi_2'\\ 0 \\ 0
   \end{array} \right ), \]
   the mixed term is equal to:
   \be
   \langle\bar{\bxi}_{1},-2(\bV_{14}-\bV_{24})\bxi'_{1}\rangle +
   \langle\bar{\bxi}_{1},(\bV_{13}-\bV_{23}+\bV_{24}-\bV_{14})\bxi'_{2}\rangle.
   \label{eq:A18}
   \ee
   Similarly the prime term is equal to:
   \[ \langle (\bxi'_1,\bxi'_2),\left (\begin{array}{rr}
	   8\bV(0)-8\bV_{24} & 2\bV_{23}-2\bV_{24}-2\bV_{34}+2\bV(0) \\
   \bV_{23}-2\bV_{24}-2\bV_{34}+2\bV(0) & 2\bV(0)-2\bV_{34}
   \end{array} \right )(\bxi'_1,\bxi'_2)^{T}\rangle\]
   \be
   +\langle(\bxi'_1,\bxi'_2),
    \left (\begin{array}{rr}
	   2\bV_{12}-2\bV(0)+4(\bV_{24}-\bV_{14}) & \bV_{13}-\bV_{23}+\bV_{24}-\bV_{14}\\
    \bV_{13}-\bV_{23}+\bV_{24}-\bV_{14} & 0
   \end{array} \right )(\bxi'_1,\bxi'_2)^{T} \rangle. \label{eq:A19}
   \ee
   The first matrix of (\ref{eq:A19}) can be expressed
   as the product:
   \[\left ( \begin{array}{rrr}
   2 & 0 & 0 \\0 & 1 & 0
   \end{array}\right ) \left (\begin{array}{rrr}
   1 & 0 & -1 \\
   0 & 1 & -1 \\
   0 & 0 & 1
   \end{array}\right ) \left ( \begin{array}{rrr}
   \bV(0) & \bV_{23} & \bV_{24} \\\bV_{23} & \bV(0) & \bV_{34} \\
   \bV_{24} & \bV_{34} & \bV(0) \end{array}\right )
   \left (\begin{array}{rrr}
   1 & 0 & 0 \\
   0 & 1 & 0 \\
   -1 & -1 & 1
   \end{array}\right )\left (\begin{array}{rr}
   2 & 0  \\
   0 & 1  \\
   0 & 0
   \end{array}\right ), \]
   hence is positive definite and bounded from below by a positive constant
   $\mu_{1}(C_1,C_2)$ times
   $r_{14}^{\zeta}|(\bxi'_1,\bxi'_2)|^{2}$. It follows that:
   \[
   Q_4(\bxi_1,\bxi_2,\bxi_3) \geq \bar{c}\rho^{\zeta}|\bar{\bxi}_1|^{2} +
   \mu_{1}(C_1,C_2)(|\bxi'_{1}|^{2} + |\bxi'_2|^{2})r_{14}^{\zeta}
   -\bar{c}\rho r_{14}^{\zeta -1}|\bar{\bxi}_{1}|(|\bxi'_1| + |\bxi'_{2}|) \]
   \[ -\bar{c}(\rho^{\zeta} + \rho r_{14}^{\zeta -1})|\bxi'_1|^{2}
   -\bar{c}\rho (r_{14}^{\zeta-1} + r_{13}^{\zeta -1})(|\bxi'_1|\cdot |\bxi'_2|)\]
   \be
   \geq \bar{c}\rho^{\zeta}|\bar{\bxi}_1|^{2} +
   \mu_{1}(C_1,C_2)(|\bxi'_{1}|^{2} + |\bxi'_{2}|^{2})r_{14}^{\zeta}
   \geq \bar{c} \rho^{\zeta}, \label{eq:A20}
   \ee
   where the mixed term is handled as for $N=3$ with Young's inequality
   and $C_0$ is chosen large enough for given $C_1$ and $C_{2}$.

   \noindent We now consider I 2.1 and I 2.2. Decompose
   $\{(\bxi_1,\bxi_2,\bxi_3,-(\bxi_1+\bxi_2 +\bxi_3)\}$
   into the orthogonal sum of
   $\{(\bar{\bxi}_1,\bar{\bxi}_2,-(\bar{\bxi}_1 +\bar{\bxi}_2), 0) \}$ and
   $\{(\bxi'_1,\bxi'_1,\bxi'_1,-3\bxi'_1)\}$. Then:
   \[ Q_4(\bxi_1,\bxi_2,\bxi_3) = \langle(\bar{\bxi}_1,\bar{\bxi}_2,-(\bar{\bxi}_1
   +\bar{\bxi}_2), 0),
   \bG_4(\bar{\bxi}_1,\bar{\bxi}_2,-(\bar{\bxi}_1 +\bar{\bxi}_2), 0)^{T}\rangle \]
   \[ + \langle((\bxi'_1,\bxi'_1,\bxi'_1,-3\bxi'_1),\bG_4
   (\bxi'_1,\bxi'_1,\bxi'_1,-3\bxi'_1)^{T}\rangle\]
   \be
   + 2 \langle(\bar{\bxi}_1,\bar{\bxi}_2,-(\bar{\bxi}_1 +\bar{\bxi}_2), 0),
   \bG_4(\bxi'_1,\bxi'_1,\bxi'_1,-3\bxi'_1)^{T}\rangle. \label{eq:A21}
   \ee
   Write:
   \[ \left (\begin{array}{r}
   \bar{\bxi}_1 \\\bar{\bxi}_2 \\ -(\bar{\bxi}_1 + \bar{\bxi}_2)\\ 0
   \end{array}\right ) =
   \left ( \begin{array}{rrrr}
   1 & 0 & 0 & 0 \\ 0 & 1 & 0 & 0 \\-1 & -1 & 1 & 0 \\0 & 0 & 0 & 1
   \end{array}\right )\left ( \begin{array}{r}
   \bar{\bxi}_1 \\ \bar{\bxi}_2 \\ 0\\ 0 \end{array}\right ). \]
   Then the bar term is equal to:
   \[ \langle (\bar{\bxi}_{1},\bar{\bxi}_{2}),
    \left ( \begin{array}{rr}
   2(\bV(0) -\bV_{13}) & \bV(0) + \bV_{12}-\bV_{13} -\bV_{23}\\
   \bV(0) + \bV_{12}-\bV_{13} -\bV_{23} & 2(\bV(0)-\bV_{23})
   \end{array}\right ) (\bar{\bxi}_{1},\bar{\bxi}_{2})^{T}\rangle,\]
   which is larger than:
   \be
    \bar{c}(\rho^{\zeta} |\bar{\bxi}_{1} -\bar{\bxi}_{2}|^{2} +
   r_{13}^{\zeta}|\bar{\bxi}_{1} + \bar{\bxi}_{2}|^{2}), \label{eq:A22}
   \ee
   by applying (\ref{eq:A13}) and the $N=3$ result. We express:
   \[
   (\bxi'_1,\bxi'_1,\bxi'_1,-3\bxi'_1)^{T} = \left (\begin{array}{rrrr}
   1 & 0 & 0 & 0 \\ 1 & 1 & 0 & 0 \\1 & 0 & 1 & 0 \\-3 & 0 & 0 & 1
   \end{array}\right ) \left (\begin{array}{r}
   \bxi'_{1} \\0 \\ 0 \\ 0
   \end{array}\right ), \]
   and so:
   \[ \bG_4\left (\begin{array}{r}
   \bxi'_{1} \\\bxi'_{1} \\\bxi'_{1} \\ -3\bxi'_{1}
   \end{array}\right ) = \left (\begin{array}{rrrr}
   \bV(0)+\bV_{12}+\bV_{13}-3\bV_{14} & \bV_{12} & \bV_{13} & \bV_{14} \\
   \bV_{12}+\bV(0)+\bV_{23}-3\bV_{24} & \bV(0) & \bV_{23}  & \bV_{24} \\
   \bV_{13} + \bV_{23} + \bV(0) -3\bV_{34} & \bV_{23} & \bV(0) & \bV_{34} \\
   \bV_{14} + \bV_{24} + \bV_{34} -3\bV(0) & \bV_{24} & \bV_{34} & \bV(0)
   \end{array}\right ) \left (\begin{array}{r}
   \bxi'_{1} \\0 \\ 0 \\ 0 \end{array}\right ). \]
   The mixed term is equal to:
   \[ 2(\bar{\bxi}_{1},\bar{\bxi}_{2},0,0)
   \left (\begin{array}{rrrr}
   1 & 0 & -1 & 0 \\ 0 & 1 & -1 & 0 \\0 & 0 & 1 & 0 \\0 & 0 & 0 & 1
   \end{array}\right )\left (\begin{array}{rrrr}
   \bV(0)+\bV_{12}+\bV_{13}-3\bV_{14} & \bV_{12} & \bV_{13} & \bV_{14} \\
   \bV_{12}+\bV(0)+\bV_{23}-3\bV_{24} & \bV(0) & \bV_{23}  & \bV_{24} \\
   \bV_{13} + \bV_{23} + \bV(0) -3\bV_{34} & \bV_{23} & \bV(0) & \bV_{34} \\
   \bV_{14} + \bV_{24} + \bV_{34} -3\bV(0) & \bV_{24} & \bV_{34} & \bV(0)
   \end{array}\right )
   \left (\begin{array}{r}
   \bxi'_{1} \\0 \\ 0 \\ 0 \end{array}\right ). \]
   \[ = 2(\bar{\bxi}_{1},\bar{\bxi}_{2})\left ( \begin{array}{rr}
   \bV_{12} -\bV_{23} - 3(\bV_{14} -\bV_{34}) & \bV_{12} -\bV_{23} \\
   \bV_{12} -\bV_{13} - 3(\bV_{24}-\bV_{34})  & \bV(0) - \bV_{23}
   \end{array}\right )
   \left (\begin{array}{r}
   \bxi'_{1}\\0 \end{array}\right ) \]
   \be
    = 2 \langle\bar{\bxi}_{1},(\bV_{12} -\bV_{23}
   -3(\bV_{14}-\bV_{34}))\bxi'_{1}\rangle
   + 2
   \langle\bar{\bxi}_{2},(\bV_{12}-\bV_{13}-3(\bV_{24}-\bV_{34}))\bxi'_{1}\rangle,
   \label{eq:A23}
   \ee
   which can be written as:
   \be
   = 2 \langle \bar{\bxi}_{1} +\bar{\bxi}_{2},
   (\bV_{12}-\bV_{23}-3(\bV_{14}-\bV_{34}))\bxi'_{1}\rangle
   + 2 \langle \bar{\bxi}_{2},
   ((\bV_{23}-\bV_{13})-3(\bV_{24}-\bV_{14}))\bxi'_{1}\rangle. \label{eq:A24}
   \ee
   It follows that the mixed term is bounded by:
   \[ \bar{c}|\bar{\bxi}_{1} +\bar{\bxi}_{2}|\cdot |\bxi'_{1}|r_{13}(
   \max(r_{12}^{\zeta -1},r_{13}^{\zeta -1}) +r_{14}^{\zeta -1} \mu(C_{2})) \]
   \[ + \bar{c}|\bar{\bxi}_{2}|r_{12}r_{13}^{\zeta -1}|\bxi'_{1}|+
   \bar{c}|\bxi'_{1}|\cdot |\bar{\bxi}_{2}|r_{12}r_{14}^{\zeta -1}. \]

   \noindent The prime term is equal to:
   \begin{eqnarray}
   & &  \langle\bxi'_{1}, (12 \bV(0) +
   2\bV_{12}+2\bV_{13}+2\bV_{23}-6\bV_{14}-6\bV_{24}-6\bV_{34})\bxi'_{1}\rangle
   \nonumber \\
   & = & \langle\bxi'_{1},18(\bV(0)-\bV_{14})\bxi'_{1}\rangle
    +
    \langle\bxi'_{1},(2(\bV_{12}-\bV(0))+
   2(\bV_{13}-\bV(0)) \nonumber \\
   & + & 2(\bV_{23}-\bV(0))+6(\bV_{24}-\bV_{34})
   -12 (\bV_{24}-\bV_{14}) )\bxi'_{1}\rangle\nonumber \\
   & \geq & \bar{c}r_{14}^{\zeta}|\bxi'_{1}|^{2} - \bar{c}
   (r_{12}^{\zeta} + r_{13}^{\zeta}+r_{23}^{\zeta})|\bxi'_{1}|^{2}
   -\mu(C_{2})(r_{23}r_{24}^{\zeta -1} + r_{12}r_{24}^{\zeta -1})|\bxi'_{1}|^{2}
   \nonumber \\
   & = & \bar{c}r_{14}^{\zeta}|\bxi'_{1}|^{2}(1 - \mu(C_{2})C_{1}^{-\zeta}
   -\mu(C_{2})C_{1}^{-1})
   \geq \bar{c}r_{14}^{\zeta}|\bxi'_{1}|^{2}, \nonumber
   \end{eqnarray}
   if $C_{1}$ is chosen large enough for given $C_{2}$.  In case of I 2.1,
   the mixed terms involving $r_{14}^{\zeta -1}$ can be controlled by a Young's
   inequality
   as in $N=3$, using $C_{1}$ sufficiently large. The terms
   $r_{12}r_{13}^{\zeta -1}|\bxi'_{1}|\cdot |\bar{\bxi}_{2}|$ and
   $r_{13}\max\{r_{12}^{\zeta -1},r_{13}^{\zeta -1}\}|\bxi'_{1}|\cdot
   |\bar{\bxi}_{1} + \bar{\bxi}_{2}|$ can be estimated by $(C'_{0})^p
   r_{12}^{\zeta}
   = r_{12}^{\zeta/2}\cdot (C'_{0})^p r_{12}^{\zeta/2}$,
   $\left(p=\max\{\zeta,1\}\right)$,
   times the $\bxi$ bar or prime factors, then using again Young's inequality,
   thanks to the relatively large coefficient $r_{14}^{\zeta}$ in front
   of $|\bxi'_{1}|^{2}$.  In other words, we use $C_{1}$ being much larger than
   any chosen $C'_{0}$. Observe that $|\bar{\bxi}_2|^2\leq
   {{1}\over{2}}|\bar{\bxi}_2-\bar{\bxi}_1|^2
   +{{1}\over{2}}|\bar{\bxi}_2+\bar{\bxi}_1|^2$, so that the mixed terms are again
   controlled by the
   prime and bar terms. In case of I 2.2, we make $C'_{0}$ itself large to
   control the term $r_{12}r_{13}^{\zeta -1}|\bxi'_{1}|\cdot |\bar{\bxi}_{2}|$.
   The
   other terms involving $r_{14}$ are standard and controlled by large $C_{1}$.
   Note that if $\zeta \in (0,1]$
   \begin{eqnarray}
   r_{13}\max\{r_{12}^{\zeta -1}, r_{13}^{\zeta -1}\} & = &
   r_{13}^{\zeta/2}r_{12}^{\zeta/2}
   \left({{r_{13}}\over{r_{12}}}\right)^{1-{{\zeta}\over{2}}} \cr
		  \, & \leq &
   (C_0')^{1-{{\zeta}\over{2}}}r_{13}^{\zeta/2}r_{12}^{\zeta/2}
   \nonumber
   \end{eqnarray}
   Thus when multiplied to $|\bar{\bxi}_{1} +\bar{\bxi}_{2}|\cdot |\bxi'_{1}|$ it
   is bounded by
   \[ {{\theta}\over{2}}r_{13}^{\zeta}|\bar{\bxi}_{1}+\bar{\bxi}_{2}|^{2}
	       +{{(C_0')^{2-\zeta}}\over{2\theta}}r_{12}^{\zeta}|\bxi'_{1}|^{2}
   \leq {{\theta}\over{2}}r_{13}^{\zeta}|\bar{\bxi}_{1}
	   +\bar{\bxi}_{2}|^{2} +{{\theta}\over{2}}r_{14}^{\zeta}|\bxi'_{1}|^{2},
    \]
   with $C_1$ much larger than chosen $C_0'$. If $\zeta \in (1,2)$, $r_{13}
   \max\{r_{12}^{\zeta -1}, r_{13}^{\zeta -1}\}=r_{13}^{\zeta}$, and its product
   with
   $|\bar{\bxi}_{1} +\bar{\bxi}_{2}|\cdot |\bxi'_{1}|$ is bounded by
   ${{\theta}\over{2}}r_{13}^{\zeta}|\bar{\bxi}_{1} +\bar{\bxi}_{2}|^{2} +
   {{r_{13}^\zeta}\over{2\theta}}|\bxi'_{1}|^{2}\leq
   {{\theta}\over{2}}r_{13}^{\zeta}|\bar{\bxi}_{1} +\bar{\bxi}_{2}|^{2} +
   {{\theta}\over{2}}r_{14}^{\zeta}|\bxi'_{1}|^{2}$, since $C_1^{-\zeta}<\theta^2$
   for large $C_1$. Summarizing the above, we conclude that:
   \be
   Q_4(\bxi_1,\bxi_2,\bxi_3) \geq \bar{c}r_{14}^{\zeta}|\bxi'_{1}|^{2} +
   \bar{c}\rho^{\zeta}|\bar{\bxi}_{1}-\bar{\bxi}_{2}|^{2} + \bar{c}r_{13}^{\zeta}
   |\bar{\bxi}_{1} + \bar{\bxi}_{2}|^{2},  \label{eq:A25}
   \ee
   which, in $(\bxi_1,\bxi_2,\bxi_3)$ variables, is:
   \be
   Q_4(\bxi_1,\bxi_2,\bxi_3) \geq \bar{c}\left( \rho^{\zeta}|\bxi_1 -\bxi_2|^{2}
   +\alpha^{\zeta}|\bxi_{1}+\bxi_{2}-2\bxi_{3}|^{2} +
   \beta^{\zeta}|\bxi_{1}+\bxi_{2}+\bxi_{3}|^{2}\right ).
   \label{eq:A26}
   \ee

   \noindent Finally we consider II. The case II 1.1 is no different from I 1.1.
   Notice that for II 1.2, we have essentially two separate scales $\beta >>
   \gamma$,
   thanks to $\alpha $ and $\beta $ being on the same scale.
   Decompose $\{(\bxi_1,\bxi_{2},\bxi_{3},-(\bxi_{1} + \bxi_{2} +
   \bxi_{3}))\}$ into the orthogonal sum of
   $\{(\bar{\bxi}_{1},-\bar{\bxi}_{1},\bar{\bxi}_{3},-\bar{\bxi}_{3})\}$ and
   $\{(\bxi'_{1},\bxi'_{1},-\bxi'_{1},-\bxi'_{1})\}$.
   The bar term is:
   \[ \langle(\bar{\bxi}_{1},-\bar{\bxi}_{1},\bar{\bxi}_{3},-\bar{\bxi}_{3}),
   \bG_4
   (\bar{\bxi}_{1},-\bar{\bxi}_{1},\bar{\bxi}_{3},-\bar{\bxi}_{3})^{T}\rangle. \]
   By writing:
   \[ \left (\begin{array}{r}
   \bar{\bxi}_{1} \\ -\bar{\bxi}_{1} \\ \bar{\bxi}_{3} \\ -\bar{\bxi}_{3}
   \end{array}\right ) =
   \left (\begin{array}{rrrr}
   1 & 0 & 0 & 0 \\-1 & 1 & 0 & 0 \\0 & 0 & 1 & 0 \\
   0 & 0 & -1 & 1 \end{array} \right )
   \left (\begin{array}{r}
   \bar{\bxi}_{1} \\ 0 \\ \bar{\bxi}_{3} \\ 0
   \end{array} \right ), \]
    we simplify the bar term into:
   \be
   \langle (\bar{\bxi}_{1},\bar{\bxi}_{3}),
   \left ( \begin{array}{rr}
   2(\bV(0)-\bV_{12}) & \bV_{13} + \bV_{24} -\bV_{14}-\bV_{23} \\
   \bV_{13} + \bV_{24} -\bV_{14}-\bV_{23} & 2 (\bV(0)-\bV_{34})
   \end{array}\right )\left (\begin{array}{r}
   \bar{\bxi}_{1} \\ \bar{\bxi}_{3} \end{array} \right ) \rangle. \label{eq:A27}
   \ee
   Then the bar term is bounded as:
   \[ = 2 \langle\bar{\bxi}_{1},(\bV(0)-\bV_{12})\bar{\bxi}_{1}\rangle +
   2\langle\bar{\bxi}_{3},(\bV(0)-\bV_{34})\bar{\bxi}_{3 }\rangle+
   2\langle\bar{\bxi}_{1},(\bV_{13}-\bV_{23}+
   \bV_{24}-\bV_{14})\bar{\bxi}_{3}\rangle\]
   \[ \geq \bar{c}\rho^{\zeta}|\bar{\bxi}_{1}|^{2} +
   \bar{c}r_{34}^{\zeta}|\bar{\bxi}_{3}|^{2} - \bar{c}|\bar{\bxi}_{1}|\cdot
   |\bar{\bxi}_{3}\left|\bV_{13}-\bV_{14}-(\bV_{23} -\bV_{24})\right|\]
   \be
    \geq {\bar{c}\over 2}\rho^{\zeta}|\bar{\bxi}_{1}|^{2} +
   {\bar{c}\over 2}r_{34}^{\zeta}|\bar{\bxi}_{3}|^{2}. \label{eq:A28}
   \ee
   To obtain the last inequality we used lemma 3:
   \[ \left|\bV_{13}-\bV_{14}-(\bV_{23} -\bV_{24})\right| \leq \bar{c}
   r_{12}r_{34}r_{13}^{\zeta -2}
   = \bar{c}r_{12}^{\zeta/2}r_{34}^{\zeta/2}\frac{r_{12}^{1-\zeta/2}
   r_{34}^{1-\zeta/2}}{r_{13}^{1-\zeta/2}r_{13}^{1-\zeta/2}}\]
   \[\leq\bar{c}r_{12}^{\zeta/2}r_{34}^{\zeta/2}
   C_1^{-(2-\zeta)/2}(C_2-1)^{-(2-\zeta)/2}. \]
   The last term is small for large $C_1,C_2$ when $\zeta<2$. Applying Young's
   inequality yields the same bound as (\ref{eq:A28}).

   \noindent Next the prime term is simplified by using:
   \[ \left (\begin{array}{r}
   \bxi'_{1} \\ \bxi'_{1} \\ -\bxi'_{1} \\ -\bxi'_{1}
   \end{array}\right ) =
   \left (\begin{array}{rrrr}
   1 & 0 & 0 & 0 \\1 & 1 & 0 & 0 \\ -1 & 0 & 1 & 0 \\ -1 & 0 & 0 & 1
    \end{array} \right )
   \left (\begin{array}{r}
   \bxi'_{1} \\ 0 \\ 0 \\ 0
   \end{array} \right ). \]
   The prime term becomes:
   \[ \langle\bxi'_{1},\left( 2\bV(0) +2\bV_{12} -2
   \bV_{13}-2\bV_{14}-2\bV_{24}-2\bV_{23} + 2\bV(0) +
   2\bV_{34}\right )\bxi'_{1} \rangle \]
   \[ = \langle\bxi'_{1},\left
   (8\bV(0)-2\bV_{13}-2\bV_{14}-2\bV_{23}-2\bV_{24}\right )\bxi'_{1}\rangle \]
   \be
    -\langle\bxi'_{1},\left ( 4\bV(0)-2\bV_{12}-2\bV_{34}\right )\bxi'_{1}\rangle
   \geq \bar{c}r_{13}^{\zeta}
   |\bxi'_{1}|^{2} \label{eq:A29}
   \ee

   \noindent The mixed bar-prime term is:
   \[\left ( \begin{array}{r}
   \bar{\bxi}_{1} \\ 0 \\ \bar{\bxi}_{3} \\0 \end{array} \right )
    \left (\begin{array}{rrrr}
   1 & -1 & 0 & 0 \\0 & 1 & 0 & 0 \\0 & 0 & 1 & -1 \\
   0 & 0 & 0 & 1 \end{array} \right )
   \left (\begin{array}{rrrr}
   \bV(0)+\bV_{12}-\bV_{13}-\bV_{14} & \bV_{12} & \bV_{13} & \bV_{14} \\
   \bV_{12}+\bV(0)-\bV_{23}-\bV_{24} & \bV(0) & \bV_{23}  & \bV_{24} \\
   \bV_{13} + \bV_{23} - \bV(0) -\bV_{34} & \bV_{23} & \bV(0) & \bV_{34} \\
   \bV_{14} + \bV_{24} -\bV_{34} -\bV(0) & \bV_{24} & \bV_{34} & \bV(0)
   \end{array}\right ) \left ( \begin{array}{r}
   \bxi'_{1} \\ 0 \\ 0 \\ 0 \end{array}\right ) \]
   \[ =
   \langle\bar{\bxi}_{1},(\bV_{23}-\bV_{13}+\bV_{24}-\bV_{14})\bxi'_{1}\rangle
   + \langle\bar{\bxi}_{3},(\bV_{13}+\bV_{23}-\bV_{14}-\bV_{24})\bxi'_{1}\rangle.
   \]
   Hence the mixed term is bounded by:
   \be
   \bar{c}(r_{12}r_{13}^{\zeta -1} + r_{12} r_{14}^{\zeta -1})|\bxi'_{1}|\cdot
   |\bar{\bxi}_{1}| + \bar{c}(r_{34}r_{14}^{\zeta -1} +r_{34}r_{24}^{\zeta -1})
   |\bxi'_{1}|\cdot |\bar{\bxi}_{3}|. \label{eq:A30}
   \ee
   All the terms in (\ref{eq:A30}) can be estimated as before with Young's
   inequality, and we have:
   \[
   Q_4(\bxi_{1},\bxi_{2},\bxi_{3}) \geq \bar{c} \rho^{\zeta}|\bar{\bxi}_{1}|^{2}
   + \bar{c}r_{34}^{\zeta}|\bar{\bxi}_{3}|^{2}, \]
   which is:
   \be
   Q_4(\bxi_{1},\bxi_{2},\bxi_{3}) \geq \bar{c}\rho^{\zeta}|\bxi_1 -\bxi_2|^{2} +
   \gamma^{\zeta}|2\bxi_{3}-\bxi_{1}-\bxi_{2}|^{2} +
   \beta^{\zeta}|\bxi_{1}+\bxi_{2}|^{2}.
   \label{eq:A31}
   \ee

   \noindent Summarizing all the cases, we finish the proof of the proposition.
   $\,\,\,\,\,\Box$

   \newpage

   \section{Properties of the N-Body Convective Operator}

   We now define a sesquilinear form $h_N[\Psi_N,\Phi_N]$ for $\Psi_N,\Phi_N\in
   L^2\left(\Omega^{\otimes N}\right)$,
   by the expression
   \be h_N[\Psi_N,\Phi_N]=\int_{\Omega^{\otimes N}}
   d\bR\,\,\overline{\grad_\bR\Psi_N(\bR)}\bdot\bG_N(\bR)
   \bdot\grad_\bR\Phi_N(\bR). \lb{sesq} \ee
   and a quadratic form $h_N[\Psi_N]=h_N[\Psi_N,\Psi_N]$. We take as the form
   domain
   \begin{eqnarray}
   \, & & \bD(h_N)=\{\Psi_N\in L^2\left(\Omega^{\otimes N}\right): \Psi_N\in
   C^\infty\left(\Omega^{\otimes N}\right),
	 {\rm supp}\Psi_N\subseteq\overline{{\Omega^{\otimes N}}_k}\,\,\,\,\,{\rm
   for}\,\,\,\,\,{\rm some}\,\,\,\,\,k, \cr
   \, & & \,\,\,\,\,\,\,\,\,\,\,\,\,\,\,\,\,\,\,\,\,\,\,\,\,\,\,\,\,\,\,\,{\rm
   and}\,\,\,\,\,\Psi_N(\bR)=0\,\,\,\,\,
	   {\rm for}\,\,\,\,\,\bR\in\partial\Omega^{\otimes N}\}. \lb{formdom}
   \end{eqnarray}
   Here we made use of an increasing sequence of open subsets of $\Omega^{\otimes
   N}$ defined as
   \be {\Omega^{\otimes N}}_k=\{\bR\in \Omega^{\otimes N}:
   \rho(\bR)>{{1}\over{k}}\}. \lb{incrseq} \ee
   Clearly, this form can be expressed as
   $h_N[\Psi_N,\Phi_N]=\langle\Psi_N,\cH_N\Phi_N\rangle$ where $\cH_N$
   is the positive, symmetric differential operator
   \be \cH_N= -{{1}\over{2}}\sum_{i,j=1}^d\sum_{n,m=1}^N\,\,
	       {{\partial}\over{\partial
   x_{in}}}\left[V_{ij}(\br_n-\br_m){{\partial}\over{\partial x_{jm}}}\cdot\right]
   \lb{symop}\ee
   with $\bD(\cH_N)=\bD(h_N)$. Our basic object of interest is the self-adjoint
   (Friedrichs) extension $\oH_N$
   of $\cH_N$, which corresponds to the operator
   with Dirichlet b.c. on $\partial\Omega^{\otimes N}$.
   Note that it will
   follow from our discussion below that the same extension $\oH_N$ also arises if
   one chooses $\bD(\cH_N)=C_0^\infty
   \left(\Omega^{\otimes N}\right)$, rather than as above. The main properties of
   $\oH_N$ follow from those of the
   form $h_N$ which we now consider.

   The basic properties of the form are contained in:
   \begin{Prop}
   The sesquilinear form $h_N[\Psi_N,\Phi_N]$ enjoys the following:

   (i) $h_N$ is a nonnegative, closable form.

   (ii) For all $\Psi_N\in\bD(h_N)$ and for the same constant $C_N$ in Proposition
   2,
   \be h_N[\Psi_N]\geq C_N\int_{\Omega^{\otimes N}}
   d\bR\,\,[\rho(\bR)]^\zeta|\grad_\bR\Psi_N(\bR)|^2. \lb{1stineq} \ee

   (iii)For all $\Psi_N\in\bD(h_N)$ and for the same constant $C_N$ in Proposition
   2,
   \be h_N[\Psi_N]\geq C_N\cdot{{(d-\gamma)^2}\over{2}}\int_{\Omega^{\otimes N}}
   d\bR\,\,
   [\rho(\bR)]^{-\gamma}|\Psi_N(\bR)|^2. \lb{2ndineq} \ee

   \end{Prop}

   \noindent {\em Proof of Proposition 3}. {\em Ad (i)}: non-negativity is obvious
   from the definition Eq.(\ref{sesq}) and
   Proposition 1{\em (i)}. That $h_N$ is closable follows from \cite{Kato},
   Theorem VI.1.27 and its Corollary
   VI.1.28. {\em Ad (ii)}: This follows directly from the definition
   Eq.(\ref{sesq}) and the variational formula
   for the minimum eigenvalue of $\bG_N(\bR)$. {\em Ad (iii)}: For the proof of
   this inequality, we use the Lemma 2
   of Lewis \cite{Lew}. That lemma states that, given an open domain $\Lambda$
   with smooth boundary, then for any function
   $g\in H^2(\Lambda)$ such that $\bigtriangleup_{\bR}g(\bR)>0$ for all
   $\bR\in\Lambda$ and for any function
   $\varphi\in C_0^\infty(\Lambda)$ (i.e. $=0$ on $\partial\Lambda$), the
   inequality holds that
   \begin{eqnarray}
   \, &  & \int_\Lambda d\bR\,\,|\bigtriangleup_\bR g(\bR)||\varphi(\bR)|^2 \cr
   \, &  & \,\,\,\,\,\,\,
	   \leq 4\int_\Lambda d\bR\,\,|\bigtriangleup_\bR g(\bR)|^{-1}|\grad_\bR
   g(\bR)|^2|\grad_\bR\varphi(\bR)|^2,
   \lb{Hardeq}
   \end{eqnarray}
   This is proved by applying Green's first formula and the Cauchy-Schwartz
   inequality (see \cite{Lew}).
   Let us take for each integer $k\geq 1$ the domain $\Lambda_k={\Omega^{\otimes
   N}}_k$ defined as in Eq.(\ref{incrseq}).
   If we define $g(\bR)=[\rho(\bR)]^\zeta$, then $g\in H^2(\Lambda_k)$ for each
   $k$ (and, in fact, $g\in C^\infty(\Lambda_k)$).
   Furthermore,
   \be \bigtriangleup_\bR g(\bR)=2\zeta(d-\gamma)[\rho(\bR)]^{-\gamma}>0,
   \lb{laplace} \ee
   for $d>\gamma$ (which certainly holds if $\zeta>0$ and $d\geq 2$) and also
   \be |\grad_\bR g(\bR)|^2=2\zeta^2[\rho(\bR)]^{2\zeta-2}. \lb{sqrgrad} \ee
   If $\Psi_N\in\bD(h_N)$, then for some $k$ sufficiently large $\Psi_N\in
   C_0^\infty(\Lambda_k)$, and all the conditions
   for the inequality (\ref{Hardeq}) are satisfied. Hence, we find by substitution
   that
   \begin{eqnarray}
   \, &  & \int_{\Omega^{\otimes
   N}}d\bR\,\,[\rho(\bR)]^\zeta|\grad_\bR\Psi_N(\bR)|^2 \cr
   \, &  & \,\,\,\,\,\,\,
	   \geq {{(d-\gamma)^2}\over{2}}\int_{\Omega^{\otimes
   N}}d\bR\,\,[\rho(\bR)]^{-\gamma}|\Psi_N(\bR)|^2,
   \lb{mainineq}
   \end{eqnarray}
   whenever $\Psi_N\in\bD(h_N)$, for $\zeta>0$ and $d\geq 2$. If we now use
   together {\em (ii)} and inequality
   (\ref{mainineq}), we obtain {\em (iii)}. $\,\,\,\,\,\Box$

   \noindent Because of item {\em (i)} we may now pass to the closed form $\ch_N$
   (see \cite{Kato}, VI.1.4). Its properties
   are given in the following Proposition 4:

   \newpage

   \begin{Prop}The sesquilinear form $\ch_N[\Psi_N,\Phi_N]$ enjoys the following:

   (i) $\ch_N$ is a nonnegative, closed form.

   (ii) The domain $\bD(\ch_N)$ consists of the Hilbert space
   $H_{h_N}\left(\Omega^{\otimes N}\right)$ obtained by
   completion of $C_0^\infty\left(\Omega^{\otimes N}\right)$ in the inner product
   \be\langle\Psi_N,\Phi_N\rangle_{h_N}=\langle\Psi_N,
   \Phi_N\rangle+h_N[\Psi_N,\Phi_N]. \lb{innprod} \ee
   In particular, $H_0^1\left(\Omega^{\otimes N}\right)\subset\bD(\ch_N)$.
   Alternatively, $\Psi_N\in\bD(\ch_N)$ iff
   $\Psi_N\in L^2\left(\Omega^{\otimes N}\right)$, its 1st distributional
   derivative satisfies $h_N[\Psi_N]<\infty$,
   and $\gamma_k\left(\left.\Psi_N\right|_{{\Omega^{\otimes N}}_k}\right)=0$ for
   all integer $k\geq 1$, where $\gamma_k$
   is the trace operator from $H^1\left({\Omega^{\otimes N}}_k\right)$ into
   $L^2\left(\partial\Omega^{\otimes N}
   \bigcap{\Omega^{\otimes N}}_k\right)$.

   (iii) Both the items (ii) and (iii) of Proposition 3 hold for $\ch_N[\Psi_N]$
   and for all $\Psi_N\in\bD(\ch_N)$. Furthermore,
   \be h_N[\Psi_N]\geq C_N
   L^{-\gamma}\cdot{{(d-\gamma)^2}\over{2}}\|\Psi_N\|^2_{L^2} \lb{3rdineq} \ee
   also for all $\Psi_N\in\bD(\ch_N)$. In particular, $\ch_N$ is strictly
   positive.
   \end{Prop}

   \noindent {\em Proof of Proposition 4:} {\em (i)} is immediate.

   \noindent {\em (ii)} We first prove the statement that
   $H_0^1\left(\Omega^{\otimes N}\right)\subset\bD(\ch_N)$.

   To see this, we remark that $\bD(h_N)$ is dense in $H^1_0\left(\Omega^{\otimes
   N}\right)$ for $d\geq 2$. In fact, it is
   well-known that in a bounded domain $\Lambda$ the set of functions
   $C^\infty_0(\Lambda-\Gamma)$, i.e. functions vanishing on
   $\Gamma\subset\Lambda$ in addition to $\Lambda^c$, is dense in $H^l_0(\Lambda)$
   if $\Gamma$ is a finite union of
   submanifolds with codimension $k\geq 2l$. This follows from standard density
   theorems for Sobolev spaces: see Ch.III of
   Adams \cite{Adams} or Ch.9 of Maz'ja \cite{Maz'ja}. The Theorem 3.23 of
   \cite{Adams} states that $C^\infty_0(\Lambda
   -\Gamma)$ is dense in $H^l_0(\Lambda)$ iff $\Gamma$ is a $(2,l)$-polar set,
   when $\Lambda=\bR^D$.  However, the same result
   is true for any open domain $\Lambda$. In fact, repeating Adams' argument, if
   $C^\infty_0(\Lambda-\Gamma)$ is not dense in
   $H^l_0(\Lambda)$, then there must be a $u\in H^l_0(\Lambda)$ and an element
   $T\in H^{-l}_0(\Lambda)$,
   the Banach dual, so that $T(u)=1$ but $T(f)=0$ for all $f\in
   C^\infty_0(\Lambda-\Gamma)$. However, by \cite{Adams},
   Theorem 3.10, this $T$ can be identified with an element of ${\cal
   D}'(\Lambda)$ supported on $\Gamma$. Since
   this can further be canonically identified with an element of ${\cal
   D}'(\bR^D)$ supported on $\Gamma$, the set
   $\Gamma$ cannot be $(2,l)$-polar. The other direction is even simpler. These
   arguments go back to \cite{HL}. On the
   other hand, by Theorem 9.2.2 of \cite{Maz'ja} the set $\Gamma$ is $(2,l)$-polar
   iff its lower $H^l$-capacity vanishes,
   $\underline{{\rm Cap}}\left(\Gamma,H^l\right)=0$. A convenient sufficient
   condition for zero $H^l$-capacity is that the
   Hausdorff $(D-2l)$-dimensional measure of $\Gamma$ be finite, ${\cal
   H}^{D-2l}\left(\Gamma\right)<\infty$. See
   Proposition 7.2.3/3 and Theorem 9.4.2 of \cite{Maz'ja}. (This is essentially
   just the converse of the Frostman theorem,
   due originally to Erd\"{o}s \& Gillis \cite{EG}.) In the case considered, the
   set $\Gamma$ is of Hausdorff dimension
   $D-k$, so that ${\cal H}^{D-2l}\left(\Gamma\right)<\infty$ for $k\geq 2l$ ($=0$
   for $k>2l$). Thus, the set $\Gamma$ has
   zero $H^l$-capacity as required.  Clearly, $\bD(h_N)$ defined in the statement
   of the Proposition 3 above coincides with
   $C^\infty_0\left(\Omega^{\otimes N}-\Gamma\right)$, where the set
   $\Gamma=\{\bR\in\Omega^{\otimes N}:\br_n=\br_m,n\neq m\}$
   has codimension $=d\geq 2$. Therefore,
   taking $D=Nd$, $l=1,\,\,k=d$ and
   $\Lambda=\Omega^{\otimes N}$ we obtain the density of
   $\bD(h_N)$ in $H^1_0\left(\Omega^{\otimes N}\right)$, as claimed.

    As a consequence, for any $\Psi_N\in H^1_0\left(\Omega^{\otimes N}\right)$
   there exists a sequence of elements
    $\Psi_N^{(m)}\in \bD(h_N)$ converging in $H^1$-norm to $\Psi_N$. Next, we
   observe that
    \be h_N[\Psi_N]\leq B_N\|\Psi_N\|_{H^1}^2  \lb{upper} \ee
    for some coefficient $B_N>0$. This may be proved by using the variational
   principle for the maximum eigenvalue
    $\lambda_N^{\max}(\bR)$ of $\bG_N(\bR)$ and then the continuity in $\bR$ of
   $\lambda_N^{\max}(\bR)$ over the compact set
    $\overline{\Omega^{\otimes N}}$ to infer $\lambda_N^{\max}(\bR)\leq B_N$. This
   inequality states that the $H^1$-norm is
    stronger than the $h_N$-seminorm. Thus, convergence in $H^1$ norm of
   $\Psi_N^{(m)}\in \bD(h_N)$ to $\Psi_N\in H^1_0\left(
    \Omega^{\otimes N}\right)$ implies both that $\Psi_N^{(m)}\rightarrow\Psi_N$
   in $L^2$ and also that $h_N[\Psi_N^{(m)}
    -\Psi_N^{(n)}]\rightarrow 0$ as $m,n\rightarrow\infty$. Comparing with
   \cite{Kato},Section VI.1.3 we see that this means
    precisely that $\Psi_N\in\bD(\ch_N)$. Therefore, $H^1_0\left(\Omega^{\otimes
   N}\right)\subset \bD(\ch_N)$. This is
    the first statement of {\em (ii)}.

    Next, we recall from \cite{Kato}, Section VI.1.3 that $\bD(\ch_N)$ is
   characterized as the Hilbert space obtained by
    completion of $\bD(h_N)$ in the inner-product (\ref{innprod}). Since
   $\bD(h_N)\subset C_0^\infty\left(\Omega^{\otimes N}
    \right)$, this is certainly contained in the Hilbert space defined in {\em
   (ii)} above. However, since we have
    shown that $H^1_0\left(\Omega^{\otimes N}\right)\subset \bD(\ch_N)$, the
   completions of $\bD(h_N)$ and $C_0^\infty
    \left(\Omega^{\otimes N}\right)$ are the same.

    Finally, we prove the alternative characterization of $\bD(\ch_N)$ in {\em
   (ii)}. We note by Proposition 3{\em (ii)} that
    for each $\Psi_N\in \bD(h_N)$ and for each $k$
    \be \|\Psi_N\|_{H^1\left({\Omega^{\otimes N}}_k\right)}\leq k^\zeta
   C_N^{-1}\cdot\|\Psi_N\|_{h_N}. \lb{lower} \ee
    Thus, the $H_{h_N}$-norm is stronger than the $H^1\left({\Omega^{\otimes
   N}}_k\right)$-norm on $\left.\bD(h_N)
    \right|_{{\Omega^{\otimes N}}_k}$. By definition, for each
   $\Psi_N\in\bD(\ch_N)$ there is a sequence $\Psi_N^{(m)}\in
    \bD(h_N)$ converging to $\Psi_N$ in $H_{h_N}$-norm. This sequence must also
   then converge to $\left.\Psi_N
    \right|_{{\Omega^{\otimes N}}_k}$ in $H^1\left({\Omega^{\otimes
   N}}_k\right)$-norm. Passing to the limit in (\ref{lower}),
    one then obtains its validity for all $\bD(\ch_N)$. This implies that
   $\left.\bD(\ch_N)\right|_{{\Omega^{\otimes N}}_k}
    \subset H^1\left({\Omega^{\otimes N}}_k\right)$ for each integer $k$.
   Furthermore, the trace $\gamma_k$ onto the
    codimension-1 set $(\partial\Omega^{\otimes N})\bigcap {\Omega^{\otimes N}}_k$
   is continuous from $H^1\left(
    {\Omega^{\otimes N}}_k\right)$ into $H^{1/2}\left((\partial\Omega^{\otimes
   N})\bigcap {\Omega^{\otimes N}}_k\right)$.
    Since $\Psi_N^{(m)}\in\bD(h_N)$, we see that
   $\gamma_k\left(\left.\Psi_N^{(m)}\right|_{{\Omega^{\otimes N}}_k}\right)=0$
    and, passing to the limit,
   $\gamma_k\left(\left.\Psi_N\right|_{{\Omega^{\otimes N}}_k}\right)=0$ as an
   element of
    $H^{1/2}\left((\partial\Omega^{\otimes N})\bigcap {\Omega^{\otimes
   N}}_k\right)$. That is the ``only if'' part of the
    characterization. The ``if'' part is very standard. For each $\Psi_N$ obeying
   the alternative set of conditions and $k\geq
    1$, we may define $\widetilde{\Psi}_N^{(k)}$ by extending the restriction
   $\left.\Psi_N\right|_{{\Omega^{\otimes N}}_k}$
    again to the whole of $\Omega^{\otimes N}$, defining it to be $0$ outside of
   ${\Omega^{\otimes N}}_k$. Because of the
    conditions on $\Psi_N$, the new function $\widetilde{\Psi}_N^{(k)}\in
   H^1_0\left(\Omega^{\otimes N}\right)$ for each
    $k\geq 1$. See Theorems 3.16 and 7.55 of \cite{Adams}. Thus,
   $\widetilde{\Psi}_N^{(k)}\in \bD(\ch_N)$ for all $k\geq 1$.
    However,
    \be
   \|\Psi_N-\widetilde{\Psi}_N^{(k)}\|_{h_N}=\left\|\left(1-\chi_{{\Omega^{\otimes
   N}}_k}\right)\Psi_N\right\|_{h_N}
    \lb{dens} \ee
    where $\chi_{{\Omega^{\otimes N}}_k}$ is the characteristic function of
   ${\Omega^{\otimes N}}_k$. Because
    $\|\Psi_N\|_{h_N}<\infty$ by assumption, the righthand side goes to zero by
   dominated convergence as $k\rightarrow\infty$.
    Thus, we conclude that
   $\lim_{k\rightarrow\infty}\|\Psi_N-\widetilde{\Psi}_N^{(k)}\|_{h_N}=0$, which
   implies that
    $\Psi_N\in\bD(\ch_N)$.

    For {\em (iii)}: We note that the righthand side of inequalities
   (\ref{1stineq}) and (\ref{2ndineq}) in Proposition 3
    {\em (ii)} \& {\em (iii)} are just certain weighted $H^1$-norms and
   $L^2$-norms, respectively, and both of these are
    bounded by the $h_N$-norm on $\bD(h_N)$. Thus, the argument used to extend
   inequality (\ref{lower}) from $\bD(h_N)$ to
    $\bD(\ch_N)$ applies also to extending (\ref{1stineq})-(\ref{2ndineq}). Noting
   that $\rho(\bR)\leq {\rm diam}\,\Omega=L$
    for all $\bR\in\Omega^{\otimes N}$, we derive inequality (\ref{3rdineq}) from
   (\ref{2ndineq}). $\,\,\,\,\Box$

   \vspace{.1 in}

    \noindent {\bf Remark:} The proof does not work for $\Omega={\bf T}^d$, the
   $d$-dimensional torus. In that case,
    inequality (\ref{1stineq}) of Proposition 3{\em (ii)} is still valid, where
   $\rho(\bR)=\min_{n\neq m,\bk\in\BZ^d}
    |\br_n-\br_m+L\cdot\bk|$ has now period $L$ in each direction as required.
   Unfortunately, the function $g(\bR)=
    [\rho(\bR)]^\zeta$ does not belong to $H^2\left(({\bf T}^d)^{\otimes
   N}\right)$ away from the set $\Gamma$ where
    $\rho(\bR)=0$. It has singularities also on the codimension-1 set $\Gamma'$ of
   points where $|\br_n-\br_m|=
    |\br_n^*-\br_m|$, with $\br_n^*$ a periodic image of $\br_n$. Unless the
   domain of $\bD(h_N)$ is chosen to be $=0$ on
    $\Gamma'$, these singularities would contribute a surface term in the Green's
   formula, invalidating (\ref{2ndineq}).
    However, if that condition on $\bD(h_N)$ is imposed, then the resulting closed
   form $\ch_N$ has Dirichlet b.c. on
    $\Gamma'$, which is unphysical. On the other hand, we expect that these are
   really just problems with the proof
    and that the inequality (\ref{2ndineq}) still holds with periodic b.c. Methods
   used to derive general Hardy-Sobolev
    inequalities (\cite{Maz'ja}, Ch.2) should apply.

   We now exploit the previous results to study the Friedrichs extension $\oH_N$
   of $\cH_N$. Its existence is provided by
   the First Representation Theorem of forms (\cite{Kato}, Theorem VI.2.1) which
   states that there is a unique self-adjoint
   operator $\oH_N$ whose domain $\bD(\oH_N)$ is a core for $\ch_N$ and for which
   $\ch_N[\Psi_N,\Phi_N]=\langle
   \Psi_N,\oH_N\Phi_N\rangle$ for every $\Psi_N\in\bD(\ch_N)$ and
   $\Phi_N\in\bD(\oH_N)$. We now discuss the
   essential properties of this operator that we will need later:
   \begin{Prop}
   The Friedrichs extension $\oH_N$ enjoys the following:

   {\em (i)} $\oH_N$ is strictly positive, with lower bound $\geq C_N
   L^{-\gamma}\cdot {{(d-\gamma)^2}\over{2}}$.

   {\em (ii)} The spectrum of $\oH_N$ is pure point.
   \end{Prop}

   \noindent {\em Proof of Proposition 5:} {\em Ad (i):} (\ref{3rdineq}) and
   \cite{Kato}, Theorem VI.2.6. {\em Ad (ii)}:
   We use the Corollary to Lemma 1 of Lewis \cite{Lew}. His hypothesis ${\cal H}1$
   is satisfied by the increasing sequence
   ${\Omega^{\otimes N}}_k$ for integer $k\geq 1$. His hypothesis ${\cal H}2$ is
   true with $H^m=H^1$ and $c_k=C_N\cdot
   k^{-\zeta}$ as a consequence of (\ref{lower}). Finally, his third hypothesis
   holds, with the role of his function
   $p(x)$ played by $C_N{{(d-\gamma)^2}\over{2}}\cdot[\rho(\bR)]^{-\gamma}$ and
   $\varepsilon_k=C_N{{(d-\gamma)^2}\over{2}}
   \cdot k^{-\gamma}$, by (\ref{2ndineq}). Lewis' proof exploits the Rellich lemma
   for the domain ${\Omega^{\otimes N}}_k$
   to show that the identity injection $I: H_{h_N}\left(\Omega^{\otimes
   N}\right)\rightarrow L^2\left(\Omega^{\otimes N}
   \right)$ is compact, by approximating it in norm with compact operators
   $I_k(\Psi_N)=\widetilde{\Psi}^{(k)}_N$,
   defined above. The segment property holds for ${\Omega^{\otimes N}}_k$, since
   its boundary is $C^\infty$ except for a
   finite number of corners where the two parts of its boundary, $\{\bR\in
   \Omega^{\otimes N}:\rho(\bR)=k\}$ and
   $\partial\Omega^{\otimes N}$, intersect. $\,\,\,\,\,\Box$

   \newpage

   \section{Proofs of the Main Theorems}

   We now prove the main results of the paper, using the properties of $\oH_N$
   proved in the preceding section. We start with:

   \noindent {\em Proof of Theorem 1:} By a stationary weak solution of
   (\ref{closeq}) at $\kappa=0$, we mean a sequence of
   $\Theta_N^*\in L^2\left(\Omega^{\otimes N}\right)$ indexed by $N\geq 1$, such
   that, for each $N\geq 1$ and for all
   $\Phi_N\in \bD(\oH_N)$,
   \be \langle\oH_N\Phi_N,\Theta_N^*\rangle=\langle \Phi_N,G_N^*\rangle
   \lb{weakeq*} \ee
   where for $N\geq 2$
   \be G_N^*(\bR)=\sum_n \of(\br_n)\Theta_{N-1}^*(...\widehat{\br_n}...)+
	       \sum_{{\rm pairs}\,\,\,\,\{nm\}}F(\br_n,\br_m)
	   \Theta_{N-2}^*(...\widehat{\br_n}...\widehat{\br_m}...)
   \lb{inhom} \ee
   is the inhomogeneous term of Eq.(\ref{closeq}) and
   $G_1^*(\br_1)=\overline{f}(\br_1)$. Because this quantity for $N>1$
   involves the correlations of lower order, our construction will proceed
   inductively. We may assume that $G_N^*\in
   L^2\left(\Omega^{\otimes N}\right)$ (in fact, $G_N^*\in
   H^1_0\left(\Omega^{\otimes N}\right)$ away from the
   set $\Gamma$). This statement is true for $N=1$ and, for $N\geq 2$, may be
   assumed to be true for all $M<N$ if the
   statement in Theorem 1 is taken as an induction hypothesis. Only the above
   regularity property of $G_N^*$ will be
   used in the induction step. Thus, it is enough to show that (\ref{weakeq*}) has
   a unique solution $\Theta_N$ for {\em any}
   $G_N\in L^2\left(\Omega^{\otimes N}\right)$ for each $N\geq 1$. That is, we
   must show that for each $N$
   \be \langle\oH_N\Phi_N,\Theta_N\rangle=\langle \Phi_N,G_N\rangle \lb{weakeq}
   \ee
   for all $\Phi_N\in\bD(\oH_N)$ has a unique solution $\Theta_N\in
   L^2\left(\Omega^{\otimes N}\right)$ for any chosen $G_N$.

   We shall first show that $\Theta_N\in\bD(\ch_N)$ where $\Theta_N$ is any weak
   solution of (\ref{weakeq}) with
   $G_N\in L^2$. To do so, we introduce the {\em smoothing operators}
   \be \cS_N^\en=(1+\en\oH_N)^{-1}, \lb{smop} \ee
   In terms of the resolvent operator $R(z,A)=(A-z)^{-1}$ this may be written as
   \be
   \cS_N^\en={{1}\over{\en}}R\left(-{{1}\over{\en}},\oH_N\right)=R(-1,\en\oH_N).
   \lb{resop} \ee
   These smoothing operators have the following properties: First, they are
   self-adjoint operators with $\|\cS^\en_N\|\leq 1$
   for all $\en>0$. Second, because $\oH_N$ is closed and $-{{1}\over{\en}}$ is in
   its resolvent set, it follows from the
   first equality of (\ref{resop}) that $\cS^\en_N:L^2\left(\Omega^{\otimes
   N}\right)\rightarrow \bD(\oH_N)$. This exhibits
   the ``smoothing'' property of the $\cS^\en_N$. Third, $\cS^\en_N$ for each
   $\en>0$ commutes with $\oH_N$, or, more
   correctly, $\cS^\en_N\oH_N\subset\oH_N\cS^\en_N$. Finally, because
   $\lim_{\en\rightarrow 0}\en\oH_N=0$ in the strong
   resolvent sense, it follows from the second equality of (\ref{resop}) that
   \be \lim_{\en\rightarrow 0}\|\cS_N^\en\Psi_N-\Psi_N\|_{L^2}=0 \lb{converg} \ee
   for all $\Psi_N\in L^2\left(\Omega^{\otimes N}\right)$. We now observe that, if
   $\Theta_N$ satisfies (\ref{weakeq}) for any
   $G_N$ in $L^2$, then for any $\Phi_N\in \bD(\oH_N)$,
   \begin{eqnarray}
   \ch_N[\Phi_N,\cS^\en_N\Theta_N] & = &
   \langle\oH_N\Phi_N,\cS^\en_N\Theta_N\rangle \cr
				   & = &
   \langle\cS^\en_N\oH_N\Phi_N,\Theta_N\rangle \cr
				   & = &
   \langle\oH_N\cS^\en_N\Phi_N,\Theta_N\rangle \cr
				   & = &
   \langle\cS^\en_N\Phi_N,G_N\rangle=\langle\Phi_N,\cS^\en_NG_N\rangle. \lb{smeq}
   \end{eqnarray}
   In particular, if we apply this to $\Phi_N=\cS^\en_N\Theta_N$, then we find for
   the quadratic form $\ch_N[\cS^\en_N\Theta_N]
   =\langle\cS^\en_N\Theta_N,\cS^\en_NG_N\rangle$ and, thus,
   \be \ch_N[\cS^\en_N\Theta_N]\leq \|\Theta_N\|_{L^2}\cdot\|G_N\|_{L^2} \lb{unbd}
   \ee
   uniformly in $\en>0$. Since, in addition, the form $\ch_N$ is closed and
   $s-\lim_{\en\rightarrow 0}\cS^\en_N\Theta_N
   =\Theta_N$ by Eq.(\ref{converg}), it follows from Theorem VI.1.16 of
   \cite{Kato} that $\Theta_N\in\bD(\ch_N)$.

   In that case, for any $\Phi_N\in \bD(\oH_N)$, the equation (\ref{weakeq}) may
   be rewritten
   \be \ch_N[\Phi_N,\Theta_N]=\langle \Phi_N,G_N\rangle. \lb{2ndweakeq} \ee
   Furthermore, $\bD(\oH_N)$ is a core for $\bD(\ch_N)$ by the First
   Representation Theorem for forms: see \cite{Kato},
   Theorem VI.2.1,item {\em (ii)}. By the same Theorem VI.2.1, item {\em (iii)},
   it follows from (\ref{2ndweakeq}) that
   $\Theta_N\in \bD(\oH_N)$ and that
   \be \oH_N\Theta_N=G_N \lb{opeq} \ee
   with equality as elements of $L^2\left(\Omega^{\otimes N}\right)$. We observe,
   since $\oH_N^{-1}$ is bounded, that
   the equation (\ref{opeq}) is equivalent to
   \be \Theta_N=\oH_N^{-1}G_N \lb{invopeq} \ee
   However, for any $G_N\in L^2\left(\Omega^{\otimes N}\right)$ the righthand side
   of (\ref{invopeq}) exists, again by
   boundedness of $\oH_N^{-1}$, and it defines an element
   $\Theta_N=\oH_N^{-1}G_N\in \bD(\oH_N)$. Thus, the weak
   solution exists and is unique. $\,\,\,\,\,\Box$

   \noindent {\em Proof of Theorem 2 (i)}: The proof of existence and uniqueness
   here very closely parallels the
   previous one, but is even easier. For this reason, we will discuss only a few
   details. As in the previous case,
   we may begin by introducing a symmetric sesquilinear form,
   \be h_N^{(\kappa_p)}[\Psi_N,\Phi_N]=h_N[\Psi_N,\Phi_N]
	       +\sum_{n=1}^N\int_{\Omega^{\otimes N}} d\bR\,\,
   \overline{(-\bigtriangleup_{\br_n})^{p/2}\Psi_N(\bR)}
   \cdot(-\bigtriangleup_{\br_n})^{p/2}\Phi_N(\bR). \lb{psesq} \ee
   densely defined on either the same domain as before,
   $\bD\left(h_N^{(\kappa_p)}\right)=\bD(h_N)$, or, with
   identical results,
   $\bD\left(h_N^{(\kappa_p)}\right)=C^\infty_0\left(\Omega^{\otimes N}\right)$.
   Clearly,
   this is the same as $h_N^{(\kappa_p)}[\Psi_N,\Phi_N]
   =\langle\Psi_N,\cH_N^{(\kappa_p)}\Phi_N\rangle$, where
   $\cH_N^{(\kappa_p)}$ is the differential operator in Eq.(\ref{psingell}) with
   $\bD\left(\cH_N^{(\kappa_p)}\right)=
   \bD\left(h_N^{(\kappa_p)}\right)$. We may now consider the self-adjoint
   (Friedrichs) extensions of these
   operators, denoted $\oH_N^{(\kappa_p)}$, just as before. We may observe that
   there is a basic inequality,
   \be \ch_N^{(\kappa_p)}[\Psi_N]\geq {{\kappa_p
   A_N}\over{L^{2p}}}\|\Psi_N\|_{L^2}^2 \lb{poinc} \ee
   with some constant $A_N>0$, for all
   $\Psi_N\in\bD\left(\ch_N^{(\kappa_p)}\right)$. This plays the same role in the
   present
   proof as inequality (\ref{3rdineq}) of Proposition 4 {\em (iii)} did in the
   previous one. It is proved first for
   $h_N^{(\kappa_p)}[\Psi_N]$ with $\Psi_N\in\bD\left(h_N^{(\kappa_p)}\right)$, by
   expanding the elements of
   $\bD\left(h_N^{(\kappa_p)}\right)$ in a series of eigenfunctions of the
   Dirichlet Laplacian $(-\bigtriangleup_\bR)_D$,
   which are complete in $H^p_0\left(\Omega^{\otimes N}\right)$. Then, the result
   is extended to $\ch_N^{(\kappa_p)}[\Psi_N]$
   by taking limits. Note that the inequality (\ref{poinc}), in particular,
   implies that the operator $\oH_N^{(\kappa_p)}$
   is strictly positive, with lower bound $\geq \kappa_p A_N/L^{2p}$. Therefore,
   the inverse operator $\left[\oH_N^{(\kappa_p)}
   \right]^{-1}$ is bounded, as before, and unique weak solutions
   $\Theta_N^{(\kappa_p)*}$ of the stationary equations
   are easily constructed with its aid.

   A last point which requires some explanation is the regularity
   $\Theta_N^{(\kappa_p)*}\in H^p_0\left(\Omega^{\otimes N}
   \right)$ of solutions. In fact, it follows as before that
   $\Theta_N^{(\kappa_p)*}\in \bD\left(\ch_N^{(\kappa_p)}\right)$.
   It therefore suffices to show that $\bD\left(\ch_N^{(\kappa_p)}\right)\subset
   H^p_0\left(\Omega^{\otimes N}\right)$.
   We may identify $\bD\left(\ch_N^{(\kappa_p)}\right)$ as the completion of the
   pre-Hilbert space $C^\infty_0\left(
   \Omega^{\otimes N}\right)$ with the inner product
   \be \langle\Psi_N,\Phi_N\rangle_{h_N^{(\kappa_p)}}
   =\langle\Psi_N,\Phi_N\rangle+h_N^{(\kappa_p)}[\Psi_N,\Phi_N].
   \lb{pinnprod} \ee
   See \cite{Kato}, Section VI.1.3. However, we have the elementary inequality
   \be \left[\sum_{n=1}^N \,k_n^2\right]^{p/2}\leq C_{N,p}\left[\sum_{n=1}^N
   \,(k_n^2)^{p/2}\right], \lb{elemineq} \ee
   with $C_{N,p}=N^{(p-2)/2}$ for $p\geq 2$ and $=1$ for $1\leq p\leq 2$. Using
   then the Parseval's equality for
   Fourier integrals, it follows that the norm $\|\Psi_N\|_{h_N^{(\kappa_p)}}$ is
   stronger on $C^\infty_0\left(
   \Omega^{\otimes N}\right)$ than the Sobolev norm
   \be \|\Psi_N\|_{H^p}^2\equiv
   \|\Psi_N\|^2_{L^2}+\|(-\bigtriangleup_\bR)^{p/2}\Psi_N\|^2_{L^2}. \lb{sobnorm}
   \ee
   Since $H^p_0\left(\Omega^{\otimes N}\right)$ is defined to be the completion of
   $C^\infty_0\left(\Omega^{\otimes N}\right)$
   in the norm $\|\cdot\|_{H^p}$, it follows that
   $\bD\left(\ch_N^{(\kappa_p)}\right)\subset H^p_0\left(\Omega^{\otimes N}
   \right)$, as required. $\,\,\,\,\,\Box$

   \noindent {\em Proof of Theorem 2 (ii):} To construct the weak-$L^2$ limits of
   $\Theta_N^{(\kappa_p)*}$ for
   $\kappa_p\rightarrow 0$, the main thing that is required are a priori estimates
   on the $L^2$-norms uniform
   in $\kappa_p>0$. These are provided as follows. First, we note that
   $\Theta_N^{(\kappa_p)*}\in \bD(\ch_N)$
   because $\Theta_N^{(\kappa_p)*}\in \bD\left(\oH_N^{(\kappa_p)}\right)$ and
   $\bD\left(\oH_N^{(\kappa_p)}\right)
   \subset H^p\left(\Omega^{\otimes N}\right)\subset \bD(\ch_N)$ for $p\geq 1$.
   Thus, we may apply Proposition 4 {\em (iii)},
   inequality (\ref{3rdineq}), to calculate that
   \begin{eqnarray}
   \|\Theta_N^{(\kappa_p)*}\|_{L^2}^2 & \leq & C_N' L^\gamma
   \ch_N\left[\Theta_N^{(\kappa_p)*}\right] \cr
				  \,  & \leq & C_N' L^\gamma
   \ch_N^{(\kappa_p)}\left[\Theta_N^{(\kappa_p)*}\right] \cr
				  \,  &  =   & C_N' L^\gamma \langle
   \Theta_N^{(\kappa_p)*},G^{(\kappa_p)*}_N\rangle \cr
				  \,  & \leq & C_N' L^\gamma
   \|\Theta_N^{(\kappa_p)*}\|_{L^2}
   \|G^{(\kappa_p)*}_N\|_{L^2}. \lb{preL2}
   \end{eqnarray}
   with $C_N'=[C_N(d-\gamma)^2/2]^{-1}$. In other words,
   \be \|\Theta_N^{(\kappa_p)*}\|_{L^2}\leq C_N' L^\gamma
   \|G^{(\kappa_p)*}_N\|_{L^2}. \lb{1stL2ineq} \ee
   Using the expression (\ref{inhom}) for $G^{(\kappa_p)*}_N$ in terms of the
   lower-order $\Theta_{M}^{(\kappa_p)*}$,
   for $M<N$, it follows that
   \be \|G^{(\kappa_p)*}_N\|_{L^2}\leq N\cdot
   \|\overline{f}\|_{L^2(\Omega)}\|\Theta_{N-1}^{(\kappa_p)*}\|_{L^2}
   +{{N(N-1)}\over{2}}\|F\|_{L^2(\Omega\otimes\Omega)}
   \|\Theta_{N-2}^{(\kappa_p)*}\|_{L^2}. \lb{indineq} \ee
   It is then straightforward to prove inductively from (\ref{1stL2ineq}) and
   (\ref{indineq}) the main $L^2$-estimates
   \be \|G_N^{(\kappa_p)*}\|_{L^2\left(\Omega^{\otimes N}\right)}\leq
   2\cdot\left(2K_N F L^\gamma\right)^{N-1}\cdot N!
   \lb{GL2} \ee
   and
   \be \|\Theta_N^{(\kappa_p)*}\|_{L^2\left(\Omega^{\otimes N}\right)}\leq
   \left(2K_N F L^\gamma\right)^N\cdot N!
   \lb{mainL2} \ee
   where $K_N=\max_{1\leq M\leq N}C_M'$ and $F=\max\{\|\overline{f}
   \|_{L^2(\Omega)},\|F\|_{L^2(\Omega\otimes\Omega)}^{1/2}\}$.

   A further crucial estimate may be extracted from the preceding discussion.
   Using the inequality $\ch_N\left[
   \Theta_N^{(\kappa_p)*}\right]\leq
   \|\Theta_N^{(\kappa_p)*}\|_{L^2}\|G^{(\kappa_p)*}_N\|_{L^2}$ contained
   in (\ref{preL2}) and the $L^2$ bound on $G_N^{(\kappa_p)*}$, (\ref{GL2}), it
   follows that
   \be  \ch_N\left[\Theta_N^{(\kappa_p)*}\right]\leq \left[\left(2K_N F
   L^\gamma\right)^N\cdot N!\right]^2.
   \lb{hNineq} \ee
   In other words, we have a uniform bound for the $\Theta_N^{(\kappa_p)*}$ in the
   norm of the Hilbert space $H_{h_N}$:
   \be \|\Theta_N^{(\kappa_p)*}\|_{h_N}\leq 2\cdot\left(2K_N F
   L^\gamma\right)^N\cdot N! \lb{hNest} \ee
   This follows by combining estimates (\ref{mainL2}) and (\ref{hNineq}).

   We now consider the weak-$L^2$ limits of $\Theta_N^{(\kappa_p)*}$ as
   $\kappa_p\rightarrow 0$. We note first, because
   of the a priori bound (\ref{mainL2}) and weak compactness of the unit ball in
   $L^2$, that any sequence $\kappa_p^{(n)}
   \rightarrow 0$ contains a subsequence $\kappa_p^{(n')}$ such that
   $w-\lim_{n'\rightarrow\infty}
   \Theta_N^{(\kappa_p^{(n')})*}=\Theta_N^{(0)*}$ exists, and
   \be \|\Theta_N^{(0)*}\|_{L^2\left(\Omega^{\otimes N}\right)}\leq \left(2K_N F
   L^\gamma\right)^N\cdot N!
   \lb{0mainL2} \ee
   Furthermore, because of the additional a priori estimate (\ref{hNest}), we may
   extract a further subsequence
   $\kappa_p^{(n'')}$ which converges weakly in $H_{h_N}$, and the limit then
   satisfies
   \be \|\Theta_N^{(0)*}\|_{h_N}\leq 2\cdot\left(2K_N F L^\gamma\right)^N\cdot N!
   \lb{0hNest} \ee
   We wish to characterize all the possible such weak sequential limits
   $\Theta_N^{(0)*}$. If the weak limits along
   subsequences $\kappa_p^{(n')}$ are all identical, then, in fact, the weak limit
   exists and equals the unique
   subsequential limit.

   We shall show that, in fact, all of the weak subsequential limits coincide with
   $\Theta_N^*$, the unique weak
   solution of the zero-diffusivity problem. First of all, we observe that for all
   $N\geq 1$ and all $\Phi_N\in
   \bD\left(\oH_N^{(\kappa_p)}\right)$
   \be \langle\left(\oH_N+\kappa_p\sum_{n=1}^N
   (-\bigtriangleup_{\br_n})^p\right)\Phi_N,\Theta_N^{(\kappa_p)*}\rangle
						   =\langle
   \Phi_N,G_N^{(\kappa_p)*}\rangle, \lb{pweakeq} \ee
   because $\Theta_N^{(\kappa_p)*}$ is a weak solution of the
   $pth$-hyperdiffusivity equation. As a consequence of
   Theorem 2{\em (i)}, we may take $\bD\left(\cH_N^{(\kappa_p)}\right)
   =\bD\left(h_N^{(\kappa_p)}\right)=C_0^\infty\left(
   \Omega^{\otimes N}\right)$. Because
   $\bD(\cH_N^{(\kappa_p)})\subset\bD(\oH_N^{(\kappa_p)})$, (\ref{pweakeq})
   is therefore true for all $\Phi_N\in C_0^\infty\left(\Omega^{\otimes
   N}\right)$, independent of the value of $\kappa_p>0$.
   Passing to the limit along subsequence $\kappa_p^{(n')}$, we then obtain
   \be \langle\oH_N\Phi_N,\Theta_N^{(0)*}\rangle=\langle \Phi_N,G_N^{(0)*}\rangle,
   \lb{0weakeq} \ee
   for all $\Phi_N\in C_0^\infty\left(\Omega^{\otimes N}\right)$. This is not
   quite the statement that
   $\Theta_N^{(0)*}$ is a weak solution of the zero-diffusivity equation, with our
   definitions. For that to be true it is
   required that (\ref{0weakeq}) hold for all $\Phi_N\in\bD(\oH_N)$. By the same
   argument as above, $C_0^\infty\left(\Omega^{
   \otimes N}\right)$ is a dense subset of $\bD(\ch_N)$ in the Hilbert space
   $H_{h_N}$. Thus, we would like to take the
   limit in $H_{h_N}$ to obtain (\ref{0weakeq}) for all $\Phi_N\in\bD(\oH_N)$, as
   required. To do so, however, requires that
   $\Theta_N^{(0)*}\in \bD(\ch_N)$, so that we may write
   \be \ch_N\left[\Phi_N,\Theta_N^{(0)*}\right]=\langle \Phi_N,G_N^{(0)*}\rangle,
   \lb{0weakeq'} \ee
   In this form, the limit may be taken to obtain (\ref{0weakeq}) for all
   $\Phi_N\in\bD(\oH_N)$. Thus, to complete the
   proof, it is enough to show that $\Theta_N^{(0)*}\in \bD(\ch_N)$.

   To demonstrate the latter regularity of $\Theta_N^{(0)*}$, we shall use the
   second characterization of $\bD(\ch_N)$ in
   Proposition 4{\em (ii)}. We already have the estimate (\ref{0hNest}). All that
   is required in addition is to show that
   \be \gamma_k\left(\left.\Theta_N^{(0)*}\right|_{{\Omega^{\otimes
   N}}_k}\right)=0 \lb{zerotr} \ee
   for all $k\geq 1$. To obtain this, we remark that for each $k$ the identity
   injection $\iota_k:H_{h_N}\left({\Omega^{
   \otimes N}}_k\right)\rightarrow H^s\left({\Omega^{\otimes N}}_k\right)$ is
   compact for any $s<1$, because the identity
   injection from $H_{h_N}\left({\Omega^{\otimes N}}_k\right)$ to
   $H^1\left({\Omega^{\otimes N}}_k\right)$ is bounded by
   (\ref{lower}) and the identity injection $H^1\left({\Omega^{\otimes
   N}}_k\right)$ into $H^s\left({\Omega^{\otimes
   N}}_k\right)$ is compact, by the Rellich lemma. We may use the above compact
   embedding for each fixed $k$ to extract by
   a diagonal argument a further subsequence $\kappa_p^{(n''')}$ such that
   \be \lim_{n'''\rightarrow\infty}\left\|\Theta_N^{(\kappa_p^{(n''')})*}
   -\Theta_N^{(0)*}\right\|_{H^s\left({\Omega^{\otimes N}}_k\right)}=0
   \lb{stHslim} \ee
   for {\em all} $k\geq 1$. However, for each $k$, the trace $\gamma_k$ is
   continuous as a map from $H^s\left({\Omega^{
   \otimes N}}_k\right)$ into $L^2\left(\partial\Omega^{\otimes
   N}\bigcap{\Omega^{\otimes N}}_k\right)$ when $s>1/2$.
   Furthermore,
   \be
   \gamma_k\left(\left.\Theta_N^{(\kappa_p^{(n''')})*}\right|_{{\Omega^{\otimes
   N}}_k}\right)=0 \lb{pzerotr} \ee
   for all $n'''$. Thus, passing to the limit, we obtain (\ref{zerotr}).
   $\,\,\,\,\,\Box$

   \section{Concluding Remarks}

   We make here just a few remarks on some further results of our analysis and
   some outstanding problems
   for future work.

   \noindent {\em (i) Regularity of the Solutions}

   \noindent The construction above produces solutions $\Theta_N^*\in
   L^2\left(\Omega^{\otimes N}\right)$ and
   $\in H^1_0\left(\Omega^{\otimes N}\right)$ away from the singular set $\Gamma$.
   In fact, as was mentioned
   in the Introduction, it is expected that $\Theta_N^*$ are H\"{o}lder regular,
   $\Theta_N^*\in C^\gamma
   \left(\Omega^{\otimes N}\right)$. Such additional regularity of the solutions
   of the singular-elliptic equations
   may follow from Harnack inequalities \cite{Mos,Trud}.

   \noindent {\em (ii) $N$-Dependence of Spectral Gap and Invariant Measure on
   Scalar Fields}

   \noindent The Proposition 2 has only been fully proved here for $N\leq 4$.
   Assuming that it holds for general $N$, the
   question of the $N$-dependence of the constant $C_N$ appearing in its statement
   has also some importance. As we have seen,
   the solutions $\Theta_N^*$ constructed for $\kappa=0$ obey an $L^2$-bound
   \be \|\Theta_N^*\|_{L^2\left(\Omega^{\otimes N}\right)}\leq B^N\cdot N!
   \lb{BmainL2} \ee
   in which $B$ is proportional to the inverse of $\min_{N\geq 1}C_N$. If $C_N$ is
   bounded from below uniformly
   in $N$, then the above constant $B<\infty$. In that case, the correlation
   functions $\Theta_N^*$ determine a
   {\em characteristic functional} via the series
   \be \Phi^*(\psi)=\sum_{N=0}^\infty {{i^N}\over{N!}}\langle\psi^{\otimes N},
   \Theta_N^*
					     \rangle_{L^2\left(\Omega^{\otimes
   N}\right)}, \lb{chfnser} \ee
   absolutely convergent for $\|\psi\|_{L^2(\Omega)}<B^{-1}$. A measure $\mu^*$ on
   scalar fields $\theta\in L^2(\Omega)$
   such that
   \be
   \Phi^*(\psi)=\int_{L^2(\Omega)}e^{i\langle\psi,\theta\rangle}\mu^*(d\theta),
   \lb{chfnint} \ee
   is therefore uniquely determined by the correlation functions. That such a
   measure actually exists is a
   consequence of the Minlos-Sazonov theorem (see \cite{GS}, Theorem V.5.1), if it
   can be shown, for example,
   that $\Phi^*$ defined by Eq.(\ref{chfnser}) is a positive-definite, weakly
   continuous functional on $L^2(\Omega)$
   and $\Theta_2^*$ is the kernel of a trace-class operator on $L^2(\Omega)$, i.e.
   $\int_\Omega d\br\,\,
   \Theta_2^*(\br,\br)<\infty$. The latter would follow from the regularity
   discussed in {\em (i)}.

   The measure $\mu^*$ so constructed would be the natural candidate for an
   invariant measure on the scalar
   fields. Whether the dynamical equation Eq.(\ref{pseq}) itself can make sense
   for $\kappa=0$ with Dirichlet
   b.c., even in a suitable weak sense, is an unresolved issue. However, for a
   periodic domain, or $\Omega={\bf T}^d$, the
   $d$-dimensional torus, there should be a sensible theory of weak solutions to
   Eq.(\ref{pseq}) and we conjecture
   that the reconstructed measure $\mu^*$ will be invariant under realizations
   evolving according to that equation.

   \noindent {\em (iii) Time-Dependent Solutions and Relaxation to the
   Steady-State}

   \noindent From our work in this paper there follow some further results for
   time-dependent solutions of the correlation
   equations, Eq.(\ref{closeq}). In fact, by standard semigroup theory (see
   \cite{Kato}, Ch.IX), a unique solution to
   Eq.(\ref{closeq}) with initial datum $\Theta_N(0)\in L^2\left(\Omega^{\otimes
   N}\right)$ may be (inductively)
   obtained via the Riemann integrals
   \be \Theta_N(t)= e^{-t\oH_N}\Theta_N(0)+\int_0^t ds\,\,e^{-(t-s)\oH_N}G_N(s)
   \lb{timdep} \ee
   with $G_N$ given by Eq.(\ref{inhom}), in terms of the strongly continuous,
   contraction semigroups $T_N(t)=e^{-t\oH_N}$.
   We refrain from precise theorem statements here. Furthermore, because of the
   strict positivity of spectrum of $\oH_N$
   established here, the semigroup $T_N(t)$ is strictly contractive and the limit
   exists
   \be \lim_{t\rightarrow
   \infty}\|\Theta_N(t)-\Theta_N^*\|_{L^2\left(\Omega^{\otimes N}\right)}=0.
   \lb{s-lim} \ee
   Thus, the time-dependent solutions converge strongly in $L^2$ to the stationary
   solutions constructed in this
   work. All of the results on existence of solutions for $\kappa_p>0$ and their
   convergence to zero-diffusivity
   solutions for $\kappa_p\rightarrow 0$, which were proved above for stationary
   solutions, also carry over to
   the time-dependent solutions.

   \vspace{0.5in}

   \noindent {\bf Acknowledgements}

   \noindent We would like to thank G. Falkovich, K. Gawedzki, R. H. Kraichnan,
   and A. Kupiainen for useful
   correspondence on these problems. The work of J. X. was partially supported by
   NSF grant
   DMS-9302830.


\begin{thebibliography}{99}
   \bibitem[1]{Kr68}R. H. Kraichnan, ``Small-scale structure of a scalar field
   convected by turbulence,'' Phys. Fluids
		    {\bf 11} 945 (1968).
   \bibitem[2]{SS}B. I. Shraiman and E. D. Siggia, ``Lagrangian path integrals and
   fluctuations in random flow,''
		  Phys. Rev. E {\bf 49} 2912 (1993).
   \bibitem[3]{Maj}A. J. Majda, ``Explicit inertial range renormalization theory
   in a model for turbulent diffusion,''
		   J. Stat. Phys. {\bf 73} 515 (1993).
   \bibitem[4]{GK-L}K. Gawedzki and A. Kupiainen, ``Universality in turbulence: an
   exactly soluble model,'' to appear
		    in proceedings of the $34^{th}$ Schladming School of Nuclear
   Physics, {\em chao-dyn/9512006}.
   \bibitem[5]{GK}K. Gawedzki and A. Kupiainen, ``Anomalous scaling of the passive
   scalar,'' Phys. Rev. Lett.
		  {\bf 75} 3834 (1995).
   \bibitem[6]{BGK}D. Bernard, K. Gawedzki and A. Kupiainen, ``Anomalous scaling
   in the N-point functions of passive
		  scalar,'' preprint (1996), {\em chao-dyn/9601018}.
   \bibitem[7]{CFKL}M. Chertkov, G. Falkovich, I. Kolokolov, and V. Lebedev,
   ``Normal and anomalous scaling of the
		    fourth-order correlation function of a randomly advected
   passive scalar,'' Phys. Rev. E {\bf 52}
		    4924 (1995).
   \bibitem[8]{CF}M. Chertkov and G. Falkovich, ``Anomalous scaling of a
   white-advected passive scalar,'' submitted
		  to Phys. Rev. Lett. (1996), {\em chao-dyn/9509007}.
   \bibitem[9]{Rich}L. F. Richardson, ``Atmospheric diffusion shown on a
   distance-neighbor graph,'' Proc. Roy. Soc.
		    Lond. A {\bf 110} 709 (1926).
   \bibitem[10]{Lew}R. T. Lewis, ``Singular elliptic operators of second order
   with purely discrete spectra,''
		   Trans. Amer. Math. Soc. {\bf 271} 653 (1982).
   \bibitem[11]{HLP}G. H. Hardy, J. E. Littlewood, and G. P\'{o}lya, {\em
   Inequalities}. (Cambridge U. Pr.,
		    Cambridge, 1934).
   \bibitem[12]{ArS}N. Aronszajn and K. T. Smith, ``Theory of Bessel potentials,
   I,'' Ann. Inst. Fourier (Grenoble),
		  {\bf 11} 385 (1961).
   \bibitem[13]{AbS}M. Abramowitz and I. A. Stegun, {\em Handbook of Mathematical
   Functions}. (National Bureau
		   of Standards, Washington, D.C.,1964).
   \bibitem[14]{Kato}T. Kato, {\em Perturbation Theory for Linear Operators.}
   (Springer-Verlag, New York, 1966).
   \bibitem[15]{Adams}R. A. Adams, {\em Sobolev Spaces.} (Academic Press, New
   York, 1975).
   \bibitem[16]{Maz'ja}V. G. Maz'ja, {\em Sobolev Spaces.} (Springer-Verlag,
   Berlin, 1985).
   \bibitem[17]{HL}L. H\"{o}rmander and J. L. Lions, ``Sur la compl\'{e}tion par
   rapport \`{a} une int\'{e}grale de
		   Dirichlet,'' Math. Scand. {\bf 4} 259 (1956).
   \bibitem[18]{EG}P. Erd\"{o}s and J. Gillis, ``Note on the transfinite
   diameter,'' J. Lond. Math. Soc. {\bf 12} 185 (1937).
   \bibitem[19]{Mos}J. Moser, ``On Harnack's theorem for elliptic differential
   equations,'' Commun. Pure Appl. Math.
		    {\bf 14} 577 (1961).
   \bibitem[20]{Trud}N. S. Trudinger, ``On Harnack type inequalities and their
   application to quasilinear
		     elliptic partial differential equations,'' Commun. Pure Appl.
   Math. {\bf 20} 721 (1967).
   \bibitem[21]{GS}I. I. Gihman and A. V. Skorohod, {\em The Theory of Stochastic
   Processes, I.}
		   (Springer-Verlag, New York, 1974).





   \end{thebibliography}
   \end{document}